\documentclass[11pt,reqno]{amsart}
\usepackage{amsmath,amssymb,mathrsfs,graphicx}
\usepackage[utf8x]{inputenc}
\usepackage[T1]{fontenc}
\usepackage[USenglish,english]{babel}
\usepackage{amsmath}
\usepackage{amsfonts}
\usepackage{latexsym}
\usepackage{cite}
\usepackage{float}
\usepackage{tikz}
\usepackage{pgfplots}
\usepackage{caption}
\usepackage{booktabs} % for using \toprule, \midrule for tables 
\usepackage{subcaption}
\usepackage{mathtools}      % This stuff is so that the equations
\mathtoolsset{showonlyrefs} % only show numbers if referenced.
\usepackage[export]{adjustbox}
\usepackage{amsthm}
\usepackage{amssymb,amscd}
\usepackage{mathtools}
\usepackage{xargs}
\usepackage{graphicx}
\usepackage{color}
\usepackage{enumitem}
\usepackage{multirow}
\usepackage{lmodern}

\usepackage[text={6.5in, 8in}, centering]{geometry}
 
\usepackage{parskip}
\usepackage{microtype}

\usepackage{hyperref}

\usepackage{subcaption}

\newtheorem*{thm*}{Theorem}

\usepackage{appendix}

\theoremstyle{definition}

\title[On Lagrangian Coherent Structures in Laparoscopy]{On Lagrangian Coherent Structures in Laparoscopy}
\author[]{Sandeep Kumar}
\address[]{Department of Quantitative methods, CUNEF Universidad, Madrid, Spain.}
\email{sandeep.kumar@cunef.edu}
\author[]{Caroline Crowley}
\address[]{School of Mechanical and Material Engineering, University College Dublin, Dublin, Ireland}
\email{caroline.crowley@ucdconnect.ie}
\author[]{Mohammad Faraz Khan}
\address[]{UCD Centre of Precision Surgery, School of Medicine, University College Dublin, Dublin, Ireland}
\email{mohammad.khan@ucd.ie}
\author[]{Miguel D. Bustamante}
\address[]{School of Mathematics and Statistics, University College Dublin, Dublin, Ireland}
\email{miguel.bustamante@ucd.ie}
\author[]{Ronan Cahill}
\address[]{Department of Surgery, Mater Misericordiae University Hospital, Dublin, Ireland}
\email{ronan.cahill@ucd.ie}
\author[]{Kevin Nolan}
\address[]{School of Mechanical and Material Engineering, University College Dublin, Dublin, Ireland}
\email{kevin.nolan@ucd.ie}

\date{\today}	
\begin{document}
\newenvironment{red}{\textcolor{red}}

\maketitle
% --------------------------------------------------------------------------------
\begin{abstract}
Laparoscopy is an electrosurgical medical operation often involving an application of high-frequency alternating current to remove undesired biological tissue from the insufflated abdomen accessible through inlet and outlets trocars. One of the main byproducts in this process are the gaseous particles, called surgical smoke, which is found hazardous for both the patient and the operating room staff. The elimination of this hazardous material is an area of active research in the medical community. Thus, understanding dynamics influenced by the underlying flow inside the abdomen is crucial. In this article, we propose a computational fluid dynamics model and analyse the velocity field in an insufflated abdomen shaped domain by identifying the Lagrangian Coherent Structures (LCS) that are responsible for the transportation, mixing and accumulation of the material particles in the flow. By calculating the mixing strength we show that the regions revealed by these material curves are dependent on the angle, positions and number of the outlets and inlets. Hence, a novel utility of LCS in medical surgery is presented that can detail the dynamics of surgical smoke informing better design of effective smoke removal technologies. 
%
% We detect these regions which locate the areas 
%
%help in an effective set up 
%
%Therefore, the utility of LCS analysis can help  of extraction of these coherent structure 

\end{abstract}
%---------------------------------------------------------------------
\section{Introduction}
%--------------------------------------------------------------------
%During a laparoscpic
%\begin{itemize}
%	\item About laparoscopy: Write from Davies 1998 article. 
%	\item Risk of $CO_2$ escape: Write from Hardy 2021, Cahill 2020.
%	\item CFD modelling: Najafabadi 2021.
%	\item FTLE and LCS tool: Suara 2020, V\'etel 2009
%	\item Structure of the article.
%\end{itemize}

Minimally Invasive Surgery (MIS), or `keyhole' (laparoscopic) surgery has become increasingly popular, thanks to its effectiveness and economical advantages \cite{sauerland2006laparoscopy}. A typical laparoscopic operation involves piercing the human abdomen wall in four or five places through which, surgical instruments including a camera are passed inside the body. The hole thus created, is held open through an inlet port, i.e., the trocar, via which carbon dioxide gas is passed inside the abdomen to maintain a working space (pneumoperitoneum). In this way, with an insufflated abdomen, with more space, the surgeons can operate while viewing the telescopic image via an attached camera and light source connected to a video monitor. Some of the main benefits of laparoscopic surgery compared to conventional surgery are faster recovery and shorter stay at the hospital, reduced scarring due to small wounds, less pain and bleeding post-surgery, thus, a shorter period of short-term disability, etc. 

Laparoscopic surgery employs long slender surgical instruments that are passed through the valved trocars. These instruments cut the tissues inside the abdomen using a high-frequency alternating electrical current \cite{McCauley2010}. The gaseous byproduct of this process is called `surgical smoke' and it not only  obscures the surgeon's vision through the camera, it can be observed that smoke and gas escape through the trocar \cite{FCC}. Several studies have addressed the hazardous nature of surgical smoke to the long term health of surgical teams \cite{dalli2020gas,dalli2020laparoscopic,hardy2021aerosols,mac2022aerosol}. For instance, the gas leaks have been quantified and well-studied with high-speed Schlieren optical imaging and the latest image processing techniques \cite{CDKFN}. The ongoing COVID-19 pandemic has increased special attention on surgical access, in particular, laparoscopic surgery, as the aerosolization hazard from surgical smoke and airborne transmission of infection is indeed concerning \cite{ZBF}. 
%\textcolor{blue}{Thus, given the advantages of MIS to patients, safety of both OR staff and environment is important and to develop a protected and safe MIS, one needs to address the problem of gas leakage, or in particular, the dynamics of the gas inside the abdomen.} 
Thus, to ensure safely of the OR staff and environment, a protected and safe MIS is desired and to that end, understanding the dynamics of gas both inside and outside the abdomen becomes very important \cite{crowley2022cfd}. It is also worth pointing that surgical masks are not traditionally considered Personal Protective Equipment (PPE) but part of the infection control chain for the patient. PPE is the last resort in the hierarchy of controls and this work is motivated by a need for better engineering controls. 

%In order to address the gas leakage, it is important to understand the behaviour of the gas starting from the moment it enters inside the abdomen. 
Computational fluid dynamics (CFD) based techniques have been largely employed to describe flow dynamics but the transport of smoke particles as reliable experimental data are sometimes difficult to obtain and can be expensive and time-consuming \cite{CY}. However, recently, CFD modelling has been employed to understand the behaviour of surgical smoke generated during laparoscopy \cite{NSCTC,crowley2022cfd}. For instance, in \cite{NSCTC}, through a numerical model and simulations, smoke generation, evacuation and quantification of its composition have been described as well as compared with experimental data, while \cite{crowley2022cfd} provides novel information about the flow dynamics of gas leaks from trocars during laparoscopic surgery by studying the trajectories of particles ejected in the leaks, using a high-fidelity CFD simulation.
%based on computational fluid dynamics (CFD) have been found very helpful to study the transport and behaviour of smoke particles, for instance, recently, it is used to quantify the composition of surgical smoke generated during laparoscopy as it evolves outside the abdomen \cite{CY,NSCTC}.
%\textbf{\textcolor{blue}{About Caroline's work!}}
On the other hand, equally important and not much addressed is the action of gas inside as it enters inside the abdomen, which also influences the movement of smoke particles. In this article, we model this process using CFD techniques and analyse different setups of laparoscopic surgery involving a change in positions, angles and number of outlets and inlet. This is important because the configuration of trocars incisions varies from one procedure to another and the angle of the trocars changes throughout an operation as instruments are manipulated. Surgical smoke is typically allowed to diffuse within the pneumoperitoneum and then vented into the operating theatre airspace in the breathing zones of theatre staff. Thus, a means to eliminate it at source requires an understanding of the flow dynamics inside the abdomen.

From the velocity field calculated from the CFD simulations one can study the particle trajectories; however, the instantaneous velocity field does not always reveal the behaviour of actual trajectories as the instantaneous streamlines can diverge from actual particle trajectories very quickly. Therefore, in a Lagrangian framework, we consider the finite time Lyapunov exponent (FTLE) which is calculated from the particle trajectories and hence, can explain the integrated effect of the flow \cite{shadden2005definition}. In other words, it measures the amount by which particles separate in a given time interval or a finite time average of the maximum expansion rate for a pair of particles advected in the flow.  Indeed, from the FTLE field, which is both space and time-dependent, complex flow patterns such as vortex formation development can be detected; moreover, it can be used to find separatrices in time-dependent systems. These separatrices are called Lagrangian Coherent Structures (LCS) and can be identified in both forward and backward FTLE fields and are analogue of stable and unstable manifolds in time-independent systems, respectively \cite{haller2015lagrangian}. 
%\textit{In fact, in \cite{haller2001distinguished}, it was suggested that, heuristically, ridges of FTLE field indicate the presence of LCS, further categorized as hyperbolic, parabolic and elliptic types based on their mathematical definitions. While the hyperbolic LCS are of repelling and attracting type corresponding to the stable and unstable manifolds, respectively, together with parabolic LCS (jet-core types), they are defined as stationary curves of the averaged material shear and the elliptic LCS (vortex type) are those of the averaged strain.} 
Note that, these structures divide dynamically distinct regions in the flow and reveal geometry which is often hidden when viewing the vector field or even trajectories of the system. Hence, they often provide an effective tool in analyzing systems with general time-dependence, especially for understanding transport of particles \cite{beron2013objective}. In this work, we have employed this approach to describe the flow dynamics inside the human abdomen during laparoscopy and thus, have established another application of the LCS technique in the field of medical surgery. 

The LCS have shown their applicability in understanding the flow in blood arteries, identifying regions of circulation, and predicting sources and fate of debris in the oceanic flow \cite{vetel2009lagrangian, suara2020material,fluids1040038, giudici2021tracking}. In this work, we employ this approach to understand the vortex formation and development in a flow, and thus claim that it can be useful in identifying regions of optimum smoke evacuation. Note that among several smoke evacuation techniques, the simplest method simply vents the smoke contaminated pneumoperitoneum into the theatre environment via a valve on the side of the trocar, which is performed manually by the surgeon. A study by Cahill et al has shown that the positive pressure ventilation in theatres is insufficient to remove pollutants when theatre staff are present as they act as a barrier to airflow \cite{CDKFN}. Other technologies claim to remove smoke via the insufflation device but have been found to be deficient \cite{dalli2020gas,dalli2020laparoscopic,hardy2021aerosols,mac2022aerosol}. The ConMed AirSeal claims to establish a gas barrier via a series of high-pressure nozzles and removing smoke through flow recirculation and filtration; this trocar has no mechanical valves. Smoke is evacuated via the trocar which in many cases is also the supply of gas. Stryker offer the PneumoClear and Neptune SafeAir series of devices. Medtronic offer similar devices. Trocar based devices collect smoke at one of the trocar locations while pencil devices collect smoke as it is generated. The former’s effectiveness is limited to timescales where the smoke has diffused and effectiveness is uncertain (despite wide use). The latter collects smoke at source but requires that surgical teams switch to completely new instruments which has an economic  and workflow impact. This is essentially an impinging jet problem. This results in the formation of vortical flow structures which will be constrained by the shape of the abdomen and the angle of the inflow jet. Effective placement of smoke evacuation requires an understanding of the internal flow structure and by addressing it, this study will inform the optimum placement of smoke removal technologies and thus, will complement the results in \cite{crowley2022cfd} in providing a complete picture of the surgical smoke problem. 

The structure of the article is the following. In Section \ref{sec:ProbSetup}, we briefly describe the problem definition, experimental setup and different types of laparoscopic surgery. Section \ref{sec:MathCompModel} elaborates the CFD model, mesh generation, the relevant system of equations supplemented with suitable boundary conditions and parameters that are essential for the validity of the model. In Section \ref{sec:ResultsDiscussion}, with a brief theoretical introduction to LCS, we consider different geometrical scenarios and analyse the velocity field obtained from the CFD model. In particular, the backward LCS are calculated by computing the particle trajectories in backward times, which identify the regions with vortex formation and potential accumulation in the two-dimension domain. In order to understand the mixing strength of these regions, clustering potential, i.e., a spatial average of backward FTLE field is calculated and using it different inlet, outlet port configurations are compared.

\section{Problem definition and Experimental set-up}
\label{sec:ProbSetup}
\begin{figure}
	\centering
	\begin{subfigure}[b]{0.2\textwidth}
		\centering
		\includegraphics[width=\textwidth]{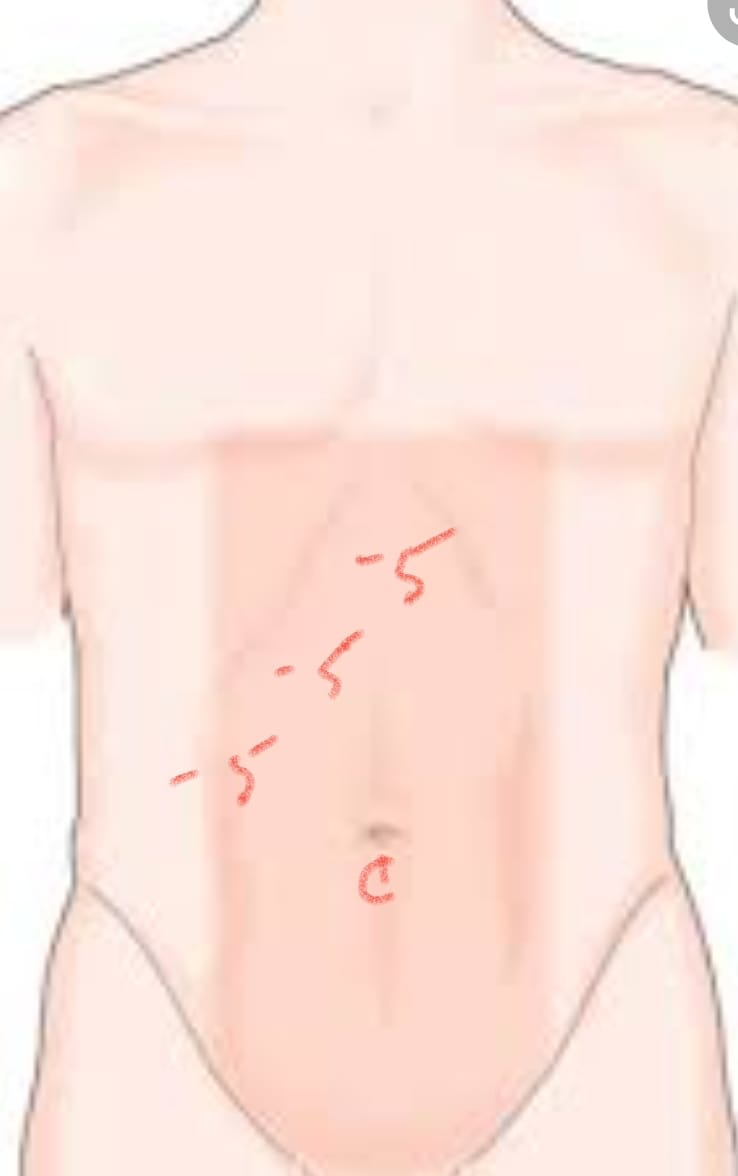}
		\caption{Case 1}
		\label{fig:Case1}
	\end{subfigure}
	\hfill
	\begin{subfigure}[b]{0.2\textwidth}
		\centering
		\includegraphics[width=\textwidth]{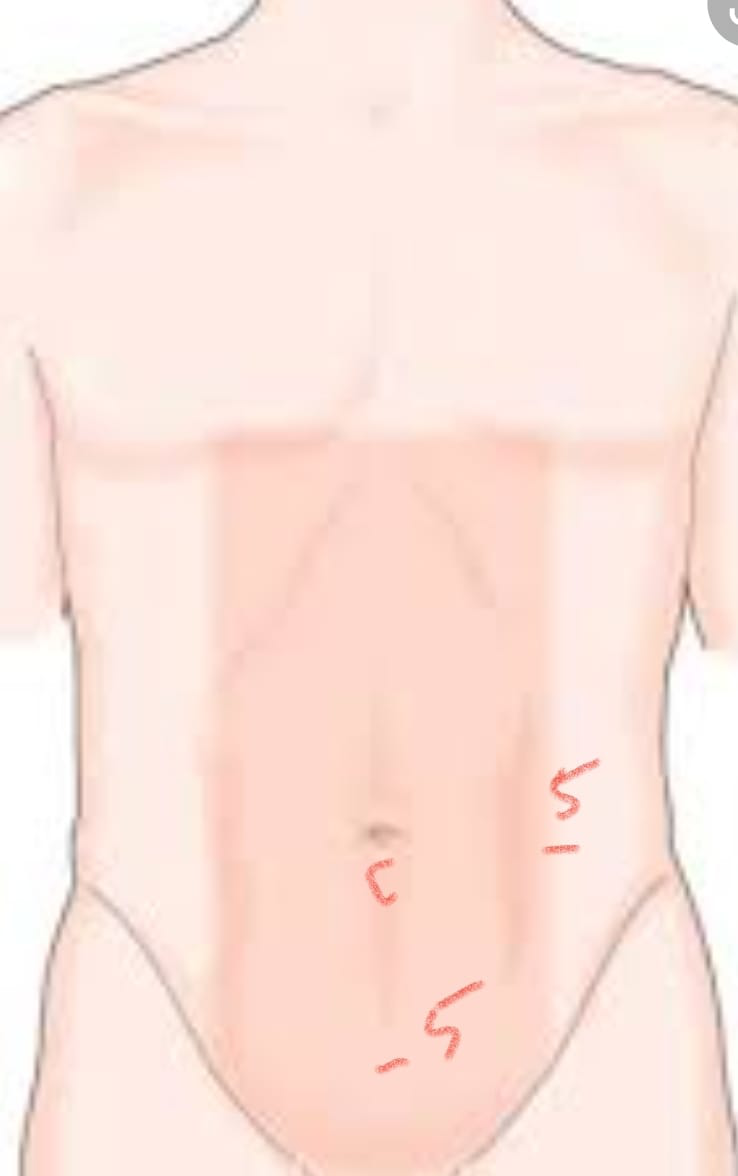}
		\caption{Case 2}
		\label{fig:Case2}
	\end{subfigure}
	\hfill		
	\begin{subfigure}[b]{0.2\textwidth}
		\centering
		\includegraphics[width=\textwidth]{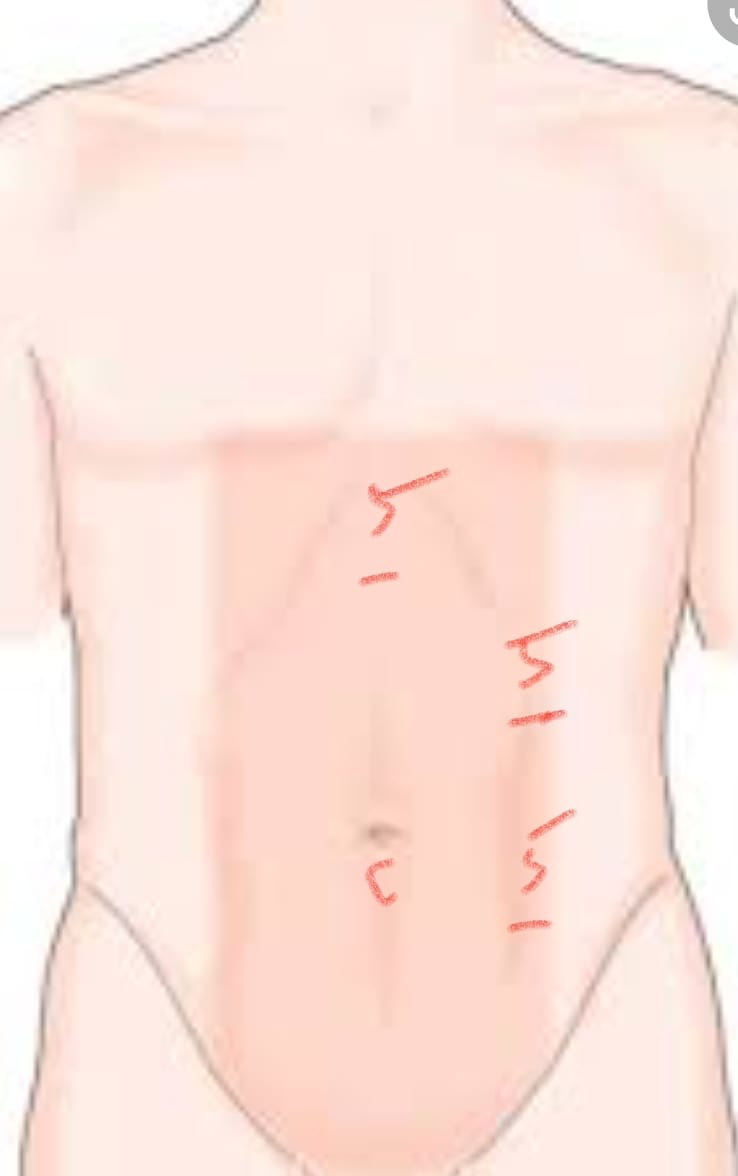}
		\caption{Case 3}
		\label{fig:Case3}
	\end{subfigure}
	\hfill	
	\begin{subfigure}[b]{0.2\textwidth}
		\centering
		\includegraphics[width=\textwidth]{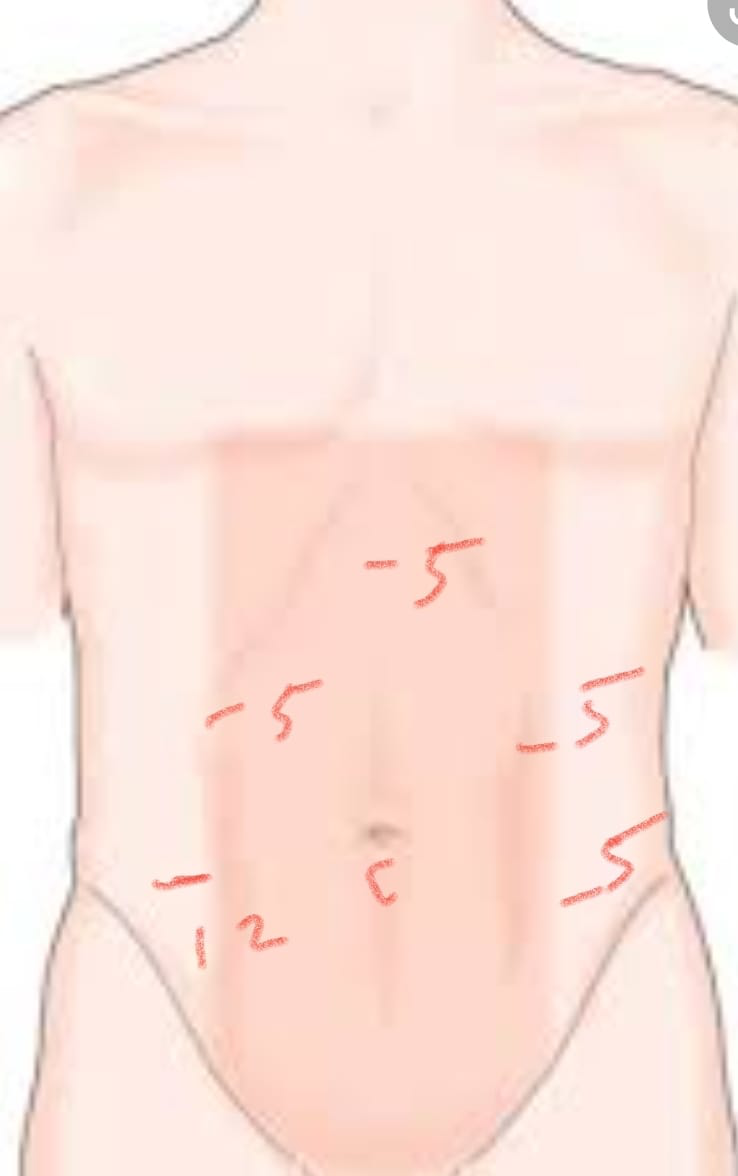}
		\caption{Case 4}
		\label{fig:Case4}
	\end{subfigure}
\caption{Four different types of laparoscopic surgery port configuration with different numbers and positions of inlet and outlets trocars. Note the camera port, which is typically also used to supply $CO_2$ is placed at the navel.}
\label{fig:Exp-set-up}				
\end{figure}

Depending on the condition of the patient and the type of the operation, the laparoscopic surgery set-up varies; nonetheless, the most common settings are the ones shown in Figure \ref{fig:Exp-set-up} which we will also consider in this work. Here, Case 1 represents laparoscopic cholecystectomy, where through the camera port, represented by `c', insufflation takes place and the rest of the three ports are used for the surgical tools, for instance, in line with `c' port is the epigastric port through which the diathermy is used for the dissection purpose. Similarly, Case 2 and 3, correspond to the setups for laparoscopic appendicectomy and laparoscopic hemicolectomy (right-sided colon surgery), respectively, where the energy device would be used from any of the two and three ports, respectively, depending on the operative stage and subsequent steps and finally, Case 4 is the left-sided colon resection or rectal surgery (see \cite{johnson1997laparoscopic} for more information). The diameter size of the ports varies from 5 mm to 12 mm, where a 12 mm port (typically only one) is used for the stapling device which is larger than the other instruments. A typical setup for Case 1 is shown in the left-hand side of Figure \ref{fig:MeshAndExpSetup}, and with this, our first goal is to model it by describing the dynamics of the flow inside the abdomen. 

%--------------------------------------------------------------------
\section{A Computational Fluid Dynamics model}
\label{sec:MathCompModel}
To model the abdomen surface in Figure \ref{fig:Exp-set-up}, we consider only the rectangular part shown in darker shaded region, and its inflated form in a simplified way can be compared with a semi-ellipsoid. The corresponding mesh for this shape has been generated using \texttt{Gmsh v4.4.1} and the measurements correspond to those of an average human body. Next, an appropriate system of mathematical equations is defined and solved numerically; the computations and analyses are performed using OpenFOAM \texttt{v9}, an open-source software library written in C++ comprising of many CFD solvers and utilities \cite{weller1998tensorial}. Several OpenFOAM utilities have been used for the simulations (e.g., mesh generation, parallelization, etc.) and for the initial postprocessing (e.g., data visualization, calculation of field numbers, etc.). The standard solvers in OpenFOAM employ finite volume discretization to solve the governing equations. 

In our case, due to the turbulent nature of the underlying phenomenon, we use the PimpleFoam solver which discretizes and solves the momentum, continuity and energy equations (Navier--Stokes equations) in each cell. PimpleFoam is a large time-step transient solver for incompressible flow and uses the PIMPLE (merged PISO-SIMPLE) algorithm. The option \texttt{adjustTimeStep} is enabled so that the maximum Courant number does not exceed 5. By setting \texttt{RASModel} variable as \texttt{kOmegaSST}, we have used the turbulence model as the $k-\omega$ SST so that the relevant transport equations for turbulent kinetic energy $k$ and the dissipation rate of turbulent kinetic energy $\epsilon$ are solved for each cell. The $k-\omega$ SST model is the most suitable choice as it is the most accurate and robust near boundaries and uses $k-\omega$ in the inner region. 

\subsection{Flow model}
We consider a 3D, Newtonian and incompressible flow and implement a computational model based on the incompressible Navier--Stokes equations with turbulence. The flow domain representing an insufflated abdomen in a simplified way is taken as a semi-ellipsoid and it has a flat base. Thus, a system of differential equations, i.e., continuity equation and the momentum equation for the conservation of mass and momentum, respectively and energy equations including a specific turbulence model is considered. More precisely, by writing the fluid velocity $\mathbf u $ and the position coordinates $\mathbf x $ in tensor form as $u_j$ and $x_j$, respectively, with direction $j$, the continuity equation is given by
\begin{equation}
\label{eq:conteq}
\frac{\partial u_j}{\partial x_j} = 0.
\end{equation}
Similarly, the momentum equation can be written as
\begin{equation}
\frac{\partial}{\partial t} (\rho u_i) + \frac{\partial}{\partial x_j} (\rho u_i u_j) = - \frac{\partial p}{\partial x_i} + \frac{\partial \tau_{ij}}{\partial x_j},
\end{equation}
where $\rho$ is the fluid density, $p$, static pressure and the shear stress $\tau_{ij}$, i.e., $j^{th}$ component of the stress acting on the faces of the fluid element perpendicular to axis $i$, is dependent on the fluid viscosity $\mu$ by
\begin{equation}
\tau_{ij} = \mu \left(\frac{\partial u_i}{\partial x_j} +\frac{\partial u_j}{\partial x_i} \right).
\end{equation}
On the other hand, the energy equations are derived using the $k-\omega$ SST turbulence model which combines the best of the $k-\omega$ model and $k-\omega$ model with a high Reynolds number. Hence, the turbulent kinetic energy $k$ and dissipation rate $\omega$ are modelled by different equations where the transport equation for $k$ is
\begin{equation}
\frac{\partial (\rho k)}{\partial t} + \frac{\partial (\rho u_ik)}{\partial x_i} = \frac{\partial }{\partial t} \left( (\mu + \sigma_k \mu_t) \frac{\partial k}{\partial x_i}\right) + \tilde P_k - \beta^\star \rho \omega k,
\end{equation}
where the left hand side corresponds to the time derivative and convection terms for $k$ and the right hand side has the diffusion, production and dissipation terms, respectively.
Similarly, the transport equation for $\omega$ is
\begin{equation}
\frac{\partial (\rho \omega)}{ \partial t } + \frac{\partial (\rho u_i \omega)}{\partial x_i} = \frac{\partial }{\partial x_i} \left( (\mu + \mu_t \sigma_\omega ) \frac{\partial \omega}{\partial x_i}\right) + \alpha \rho S^2 - \beta \rho \omega^2 + 2(1-F_1) \frac{\rho \sigma_{\omega_2}}{\omega} \frac{\partial k}{\partial x_j} \frac{\partial \omega}{\partial x_j}, 
\end{equation}
where the blending function 
$$
F_1 = \tanh\left(\left(\min \left(\max \left(\frac{\sqrt{k}}{\beta^\star
	\omega y}, \frac{500\nu}{y^2\omega}\right), \frac{4\rho \sigma_{\omega_2} k }{CD_{k\omega}y^2}\right)\right)^4\right),
$$ 
takes a value of 1 at the near wall region to activate the original equation for $\omega$ , and gradually switch to 0 moving away from the surface to activate the transformed $k-\epsilon$ equation. Here, $y$ denotes the distance to the nearest surface and 
$$
CD_{k\omega} = \max\left(\frac{2\rho\sigma_{\omega_2}}{\omega} \frac{\partial k}{\partial x_j}\frac{\partial \omega}{\partial x_j}, 10^{-10}\right).
$$
The turbulent eddy viscosity is defined as
\begin{equation}
\nu_t = \frac{a_1k}{\max(a_1\omega, SF_2)},
\end{equation}
where $S$ denotes invariant measure of strain rate and $F_2$ is the second blending function that determines the value of $\nu_t$ to be taken and is given by 
$$
F_2 = \tanh\left(\left(\max \left(2\frac{\sqrt{k}}{\beta^\star
	\omega y}, \frac{500\nu}{y^2\omega}\right), \frac{4\rho \sigma_{\omega_2} k }{CD_{k\omega}y^2}\right)^2\right).
$$ 
The values of the constants are $\beta^\star = 0.09$, $\alpha_1=5/9$, $\beta_1 = 3/40$, $\sigma_{k_1}=0.85$, $\sigma_{\omega_1}=0.5$, $\alpha_2=0.44$, $\beta_2=0.0828$, $\sigma_{k_2}=1$, $\sigma_{\omega_1}=0.856$.

\subsection{Mesh generation and CFD simulations}
The mesh has been generated using \texttt{Gmsh v4.4.1} and the semi-ellipsoid has been obtained from a semi-sphere of 0.28 m diameter dilated in $y$ direction 1.5 times. The inlet and outlets of radii  5mm are placed as shown in Figure \ref{fig:Exp-set-up} and to have a finer mesh around them, \texttt{transfinite Curve} option with a value 20 is used. The \texttt{Element size factor} is equal to 0.1. The unstructured polyhedral mesh consists of predominantly tetrahedra finite volume cells and the quality of the mesh thus generated is assessed for the original configuration with the OpenFOAM utility \texttt{checkMesh}. 
%\textcolor{blue}{The parameters for each of the four cases are shown in the table XXX.} 

\begin{figure}
	\centering
	\begin{subfigure}[b]{0.48\textwidth}
		\centering
		\includegraphics[width=\textwidth]{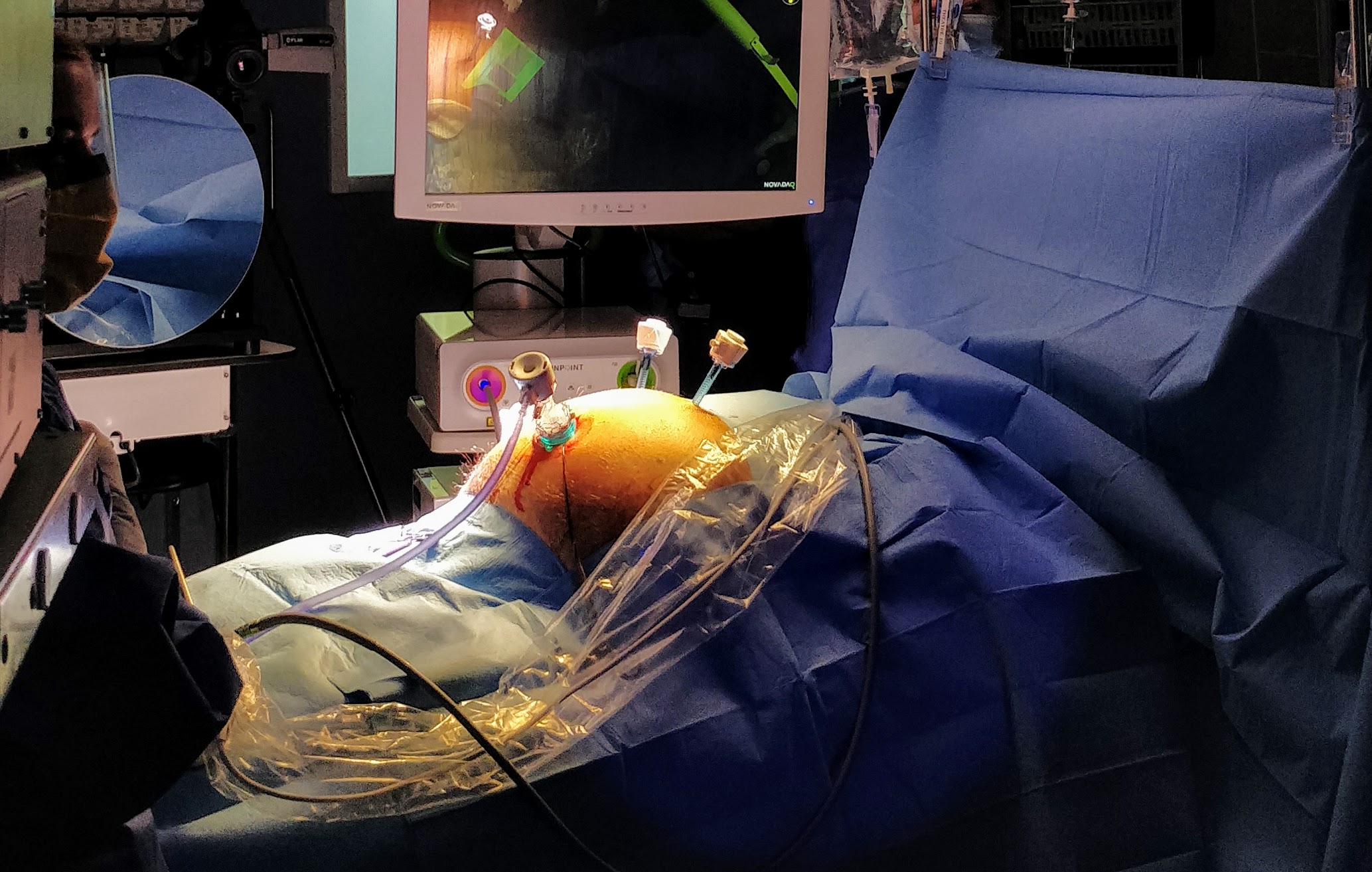}
		\caption{A laparoscopic cholecystectomy at Mater Misericordiae University Hospital, Dublin, Ireland.}
		\label{fig:ExpSetup}
	\end{subfigure}
	\hfill
	\begin{subfigure}[b]{0.5\textwidth}
		\centering
		\includegraphics[width=\textwidth]{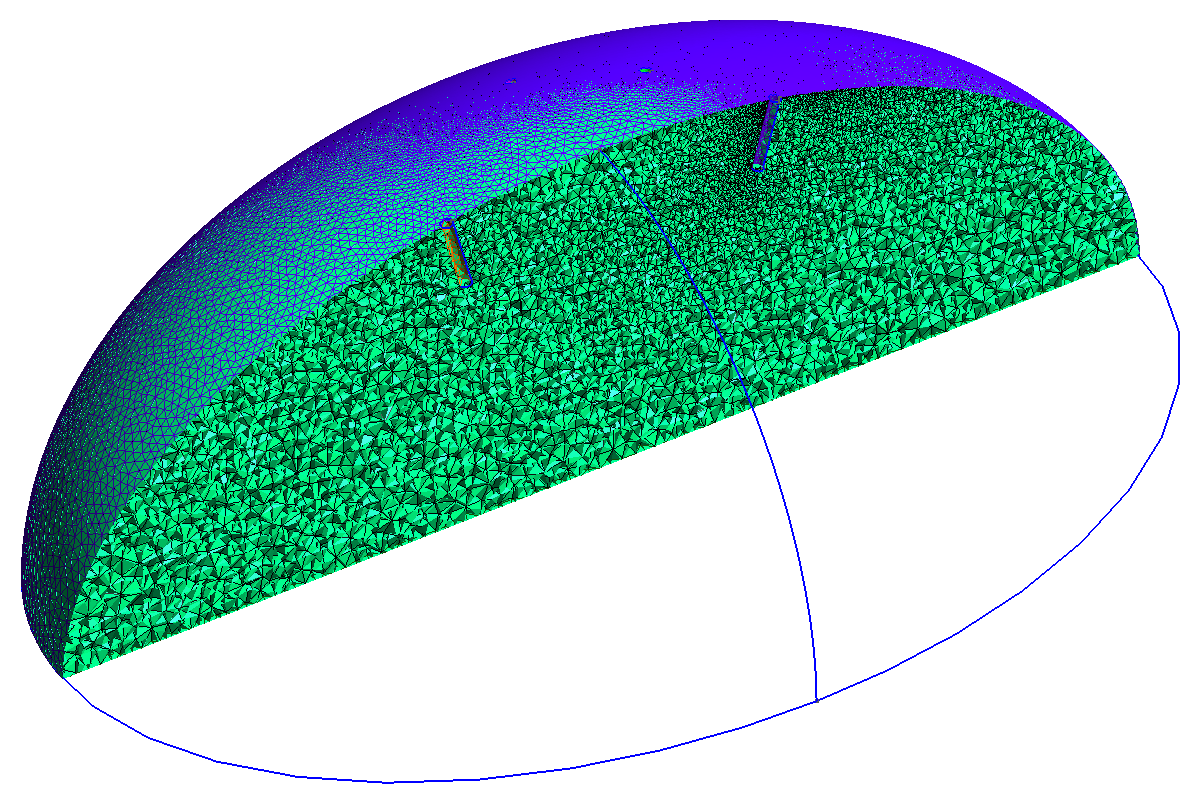}
		\caption{Mesh created in \texttt{Gmsh} for a simplified geometry.}
		\label{fig:Mesh3D}
	\end{subfigure}
\caption{}
\label{fig:MeshAndExpSetup}
\end{figure}

With this, we compute the time evolution by solving the initial boundary value problem with the PimpleFoam solver with the $k-\omega$ SST model. The numerical schemes and the boundary conditions imposed are shown in Tables \ref{tab:NumScheme3D} and \ref{tab:BC3D}, respectively. The Dirichlet boundary conditions are initialialized to zero for $\nu_t$, $\phi$, $p$ (outlets), $U$ (abdomen wall) and to 440.15 for $\omega$ (inlet) and for $k$ (inlet) turbulent intensity kinetic energy is set to 0.375. The wall functions are used for $\epsilon$ equal to 200, $k$ equal to 0, and $\omega$ (abdomen wall) equal to 440.15. We consider flow rates of volumetric type for both inlet and outlets and value for inlet flow rate $V$ is set 4.71 L/min. Using the trocar's radius $r_{in}=5mm$, we calculated the inlet velocity $U_{in} = V/\pi r_{in}^2$ equal to 1 m/s; the direction of the velocity is set as that of the inlet as shown in Figure \ref{fig:MeshAndExpSetup}.

%the direction of the velocity is set in the negative $y$ direction. 

% -------------------------------------------------
\begin{table}
	\centering
	\begin{tabular}{ |p{4cm}||p{3cm}|p{5cm} |  }
	\hline
%	\multicolumn{3}{|c|}{} \\
%	\hline
	Term & Method & OpenFOAM  \\
	\hline
	Time   & Euler    & \texttt{Euler} \\
	Gradient & Gauss  & \texttt{linear} \\
	Divergence & Gauss  & \texttt{Gauss limitedLinear 1} \\
	Laplace & Gauss & \texttt{Gauss linear corrected} \\
	Interpolation & - & \texttt{linear} \\
	Surface-normal gradient & - & \texttt{corrected}\\
	\hline
	\end{tabular}
	\caption{The numerical schemes for the CFD model.}
	\label{tab:NumScheme3D}
\end{table}
\begin{table}
	\centering
	\begin{tabular}{ |p{3cm}||p{3cm}|p{4cm} | p{4cm} | }
%		\hline
%		\multicolumn{4}{|c|}{} \\
		\hline 
		Boundary & Field & Boundary condition & OpenFOAM\\
		\hline
		Inlet & $\omega,U, \phi, \nu_t, k$ & Dirichlet & \texttt{fixedValue} \\
		Inlet & $\epsilon, \tilde{\nu},p$ & Neumann& \texttt{zeroGradient} \\
		Outlets & $\omega, p, \phi, \nu_t$ & Dirichlet & \texttt{fixedValue} \\
		Outlets & $\epsilon,k, \tilde{\nu}, \omega, U$ & Neumann& \texttt{zeroGradient} \\
		Abdomen wall & $\omega, U, \phi, \nu_t, k, \epsilon$ & Dirichlet & \texttt{fixedValue} \\
		Abdomen wall & $p, \tilde{\nu}$ & Neumann & \texttt{zeroGradient} \\		
		\hline
	\end{tabular}
	\caption{The boundary conditions for the CFD model.}
	\label{tab:BC3D}	
\end{table}

\begin{figure}
	\centering
	\begin{subfigure}[b]{0.48\textwidth}
		\centering
		\includegraphics[width=\textwidth]{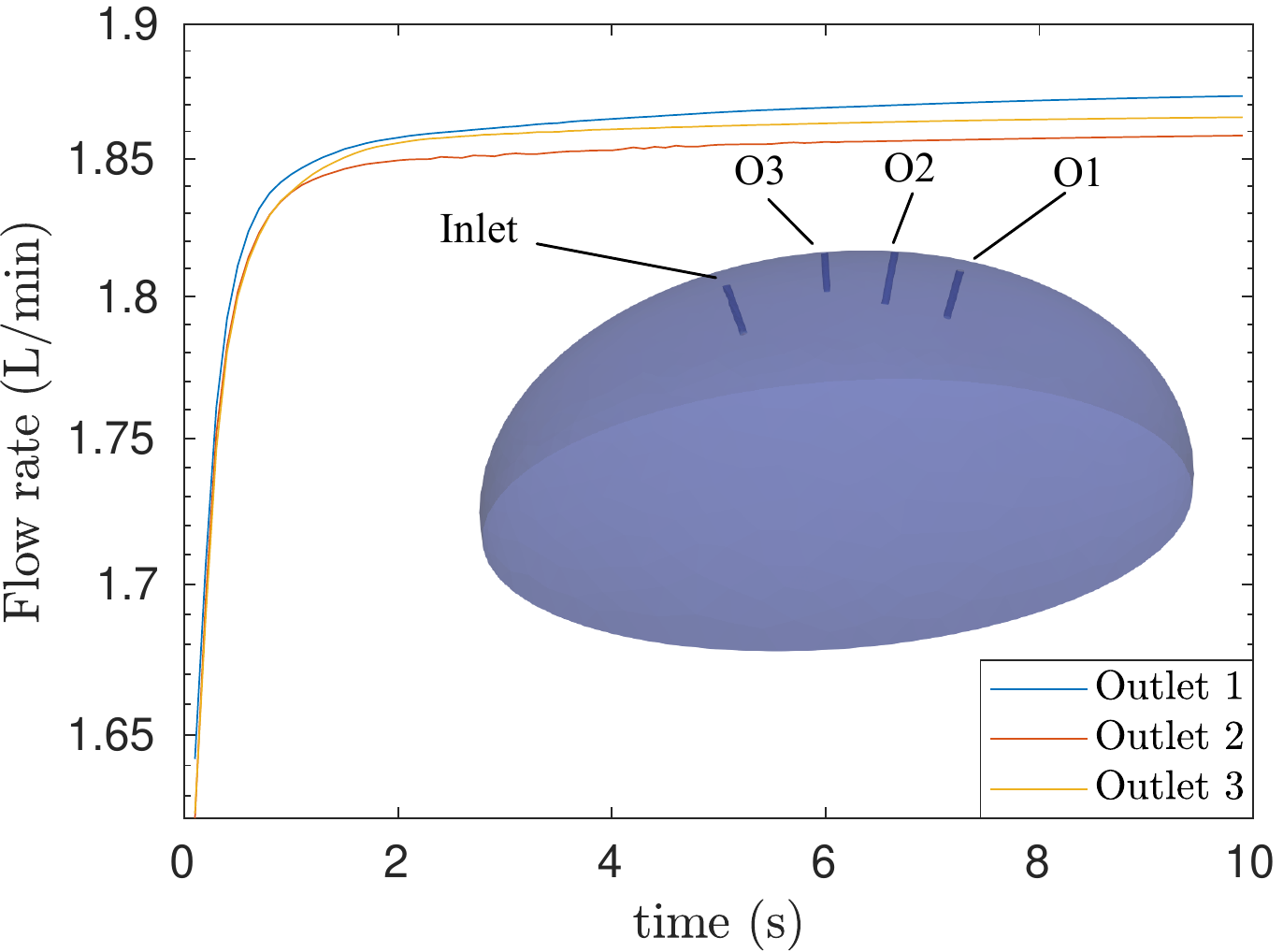}
		\caption{Case 1}
		\label{fig:FlowRate1}
	\end{subfigure}
	\hfill
	\begin{subfigure}[b]{0.488\textwidth}
		\centering
		\includegraphics[width=\textwidth]{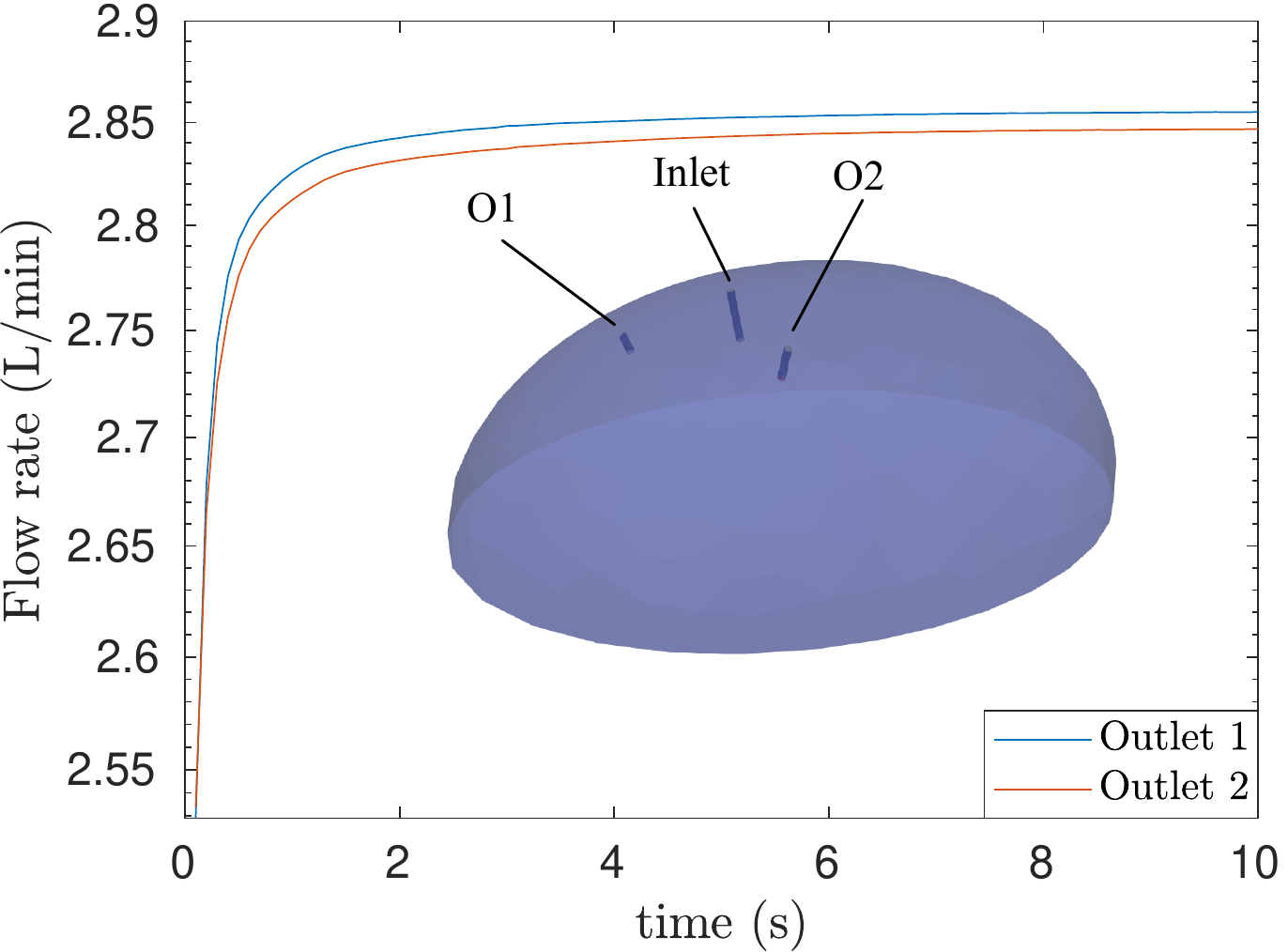}
		\caption{Case 2}
		\label{fig:FlowRate2}
	\end{subfigure}
	\hfill
	\begin{subfigure}[b]{0.488\textwidth}
		\centering
		\includegraphics[width=\textwidth]{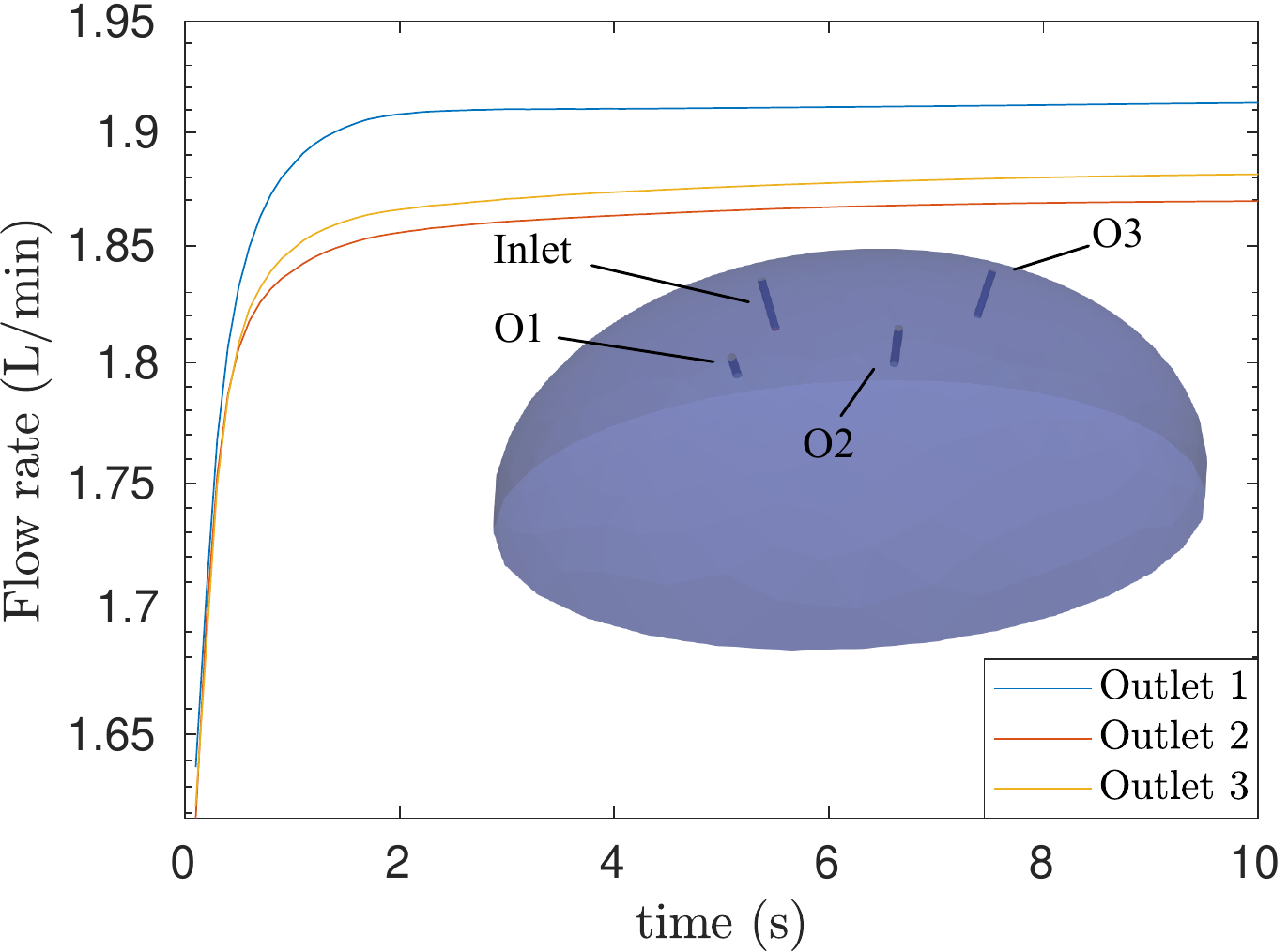}
		\caption{Case 3}
		\label{fig:FlowRate3}
	\end{subfigure}
	\begin{subfigure}[b]{0.49\textwidth}
		\centering
		\includegraphics[width=\textwidth]{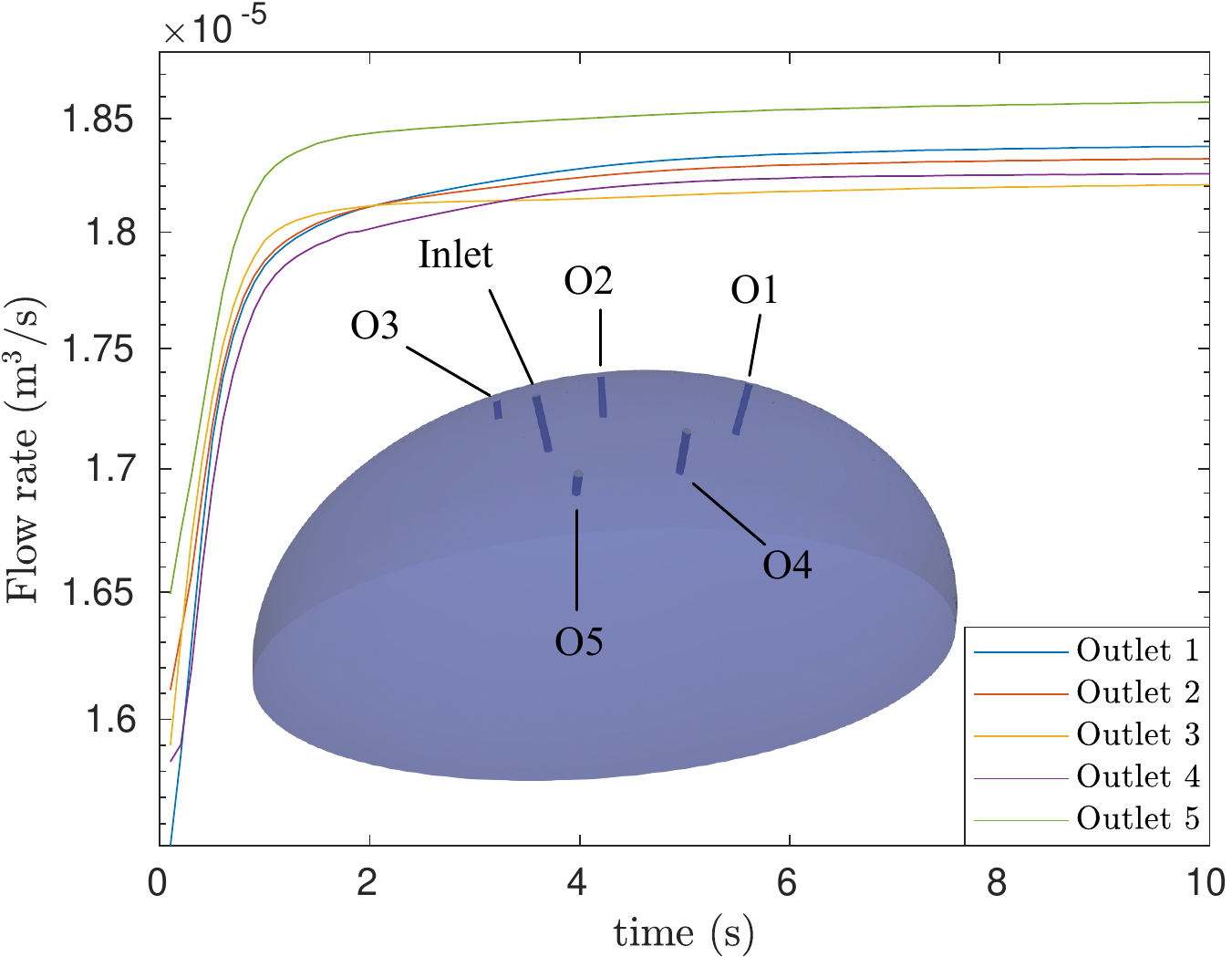}
		\caption{Case 4}
		\label{fig:FlowRate4}
	\end{subfigure}
	\caption{Flow rates for each of the four cases shown in Figure \ref{fig:Exp-set-up} plotted on a semilogarithmic scale. Also shown is the corresponding domain geometry with inlet and different outlets indicated as O1, O2, representing Outlet 1, Outlet 2, etc., respectively.  Clearly, their flow rate depends on the position and number of the outlet trocars.}  
	\label{fig:FlowRates}	
\end{figure}

%\begin{figure}	
%	\centering
%	\includegraphics[width=\textwidth]{FTLEWholeabdCase1}
%		\begin{subfigure}[b]{0.568\textwidth}
%		\centering
%		\includegraphics[width=\textwidth]{StretchlineCenterCase1}
%		\caption{}
%		\label{fig:StretchCenterCase1}
%	\end{subfigure}
%	\begin{subfigure}[b]{0.415\textwidth}
%		\centering
%		\includegraphics[width=\textwidth]{StretchlineRightCase1}
%		\caption{}
%		\label{fig:StrechRightCase1}
%	\end{subfigure}
%	\begin{subfigure}[b]{0.568\textwidth}
%		\centering
%		\includegraphics[width=\textwidth]{ShrinklineCenterCase1}
%		\caption{}
%		\label{fig:ShrinkCenterCase1}
%	\end{subfigure}
%	\begin{subfigure}[b]{0.415\textwidth}
%		\centering
%		\includegraphics[width=\textwidth]{ShrinklineRightCase1}
%		\caption{}
%		\label{fig:ShrinkRightCase1}
%	\end{subfigure}
%	\label{fig:FTLECase1}	
%	\caption{Case 1}
%\end{figure}

%\textcolor{red}{Make a table with all parameters of CFD}

\section{Results and Discussion}
\label{sec:ResultsDiscussion}
With the parameters and methodology discussed above, we have calculated the numerical evolution for a time period of 10 seconds. In particular, we compute the flow rates for each of the outlets for different cases as shown in Figure \ref{fig:FlowRates} where the flow rate is plotted against time. The different colours correspond to different outlets and their values clearly indicate that the flow rates depend on the position and number of outlets. The numerical simulations (\href{https://youtu.be/xMFBq5DtMQc}{link}) show the evolution of smoke particles that follow the eddies generated after the injection of the gas. Indeed, the movement (transportation, mixing) of these particles is influenced by the underlying flow dynamics and to understand it, one can not always rely upon the instantaneous velocity field as the instantaneous streamlines can diverge from actual particle trajectories very quickly \cite{shadden2005definition}. Furthermore, these streamlines vary when viewed from different reference frames, for instance, a region with closed streamlines in one frame can appear completely different when viewed in another frame, and they do not serve for understanding the material transport. Thus, instead of an Euler perspective, we work in a Lagrangian framework and for a given time interval, we calculate the particle trajectories and quantities that are also frame-invariant, and reveal the flow structures such as circulation, vortex formation as discussed in the following. 
\subsection{Lagrangian Coherent Structures and their application}
Given a two-dimensional velocity field $\mathbf{u}(\mathbf x, t)$ of the form
\begin{equation}
\label{eq:xprime}
\mathbf{x}^\prime = \mathbf {u} (\mathbf x, t), \ \mathbf{x}\in U \subset \mathbb{R}^2, \ t\in[t_-, t_+],
\end{equation}
the trajectories it generates are denoted by $\mathbf{x}(t; \mathbf{x}_0, t_0 )$, with a position $\mathbf x_0$ at time $t_0$. The dynamics of fluid elements can be explained with the flow map
\begin{equation}
F^t_{t_0}(\mathbf x_0) \equiv \mathbf x(t; t_0, \mathbf x_0),
\end{equation}
mapping initial positions $\mathbf x_0$ to current positions at time $t$. Using the flow map, the right Cauchy--Green strain tensor field $C_{t_0}^t(\mathbf x_0)=\nabla F_{t_0}^t(\mathbf x_0)^T \nabla F_{t_0}^t(\mathbf x_0)$ is defined which characterizes Lagrangian strain in the flow. It is symmetric and positive definite and thus, its eigenvalues $\lambda_j(\mathbf x_0)$ and eigenvectors $\xi_j(\mathbf x_0)$ satisfy 
\begin{align}
\label{eq:CauchyGreen}
	C_{t_0}^t \xi_j &= \lambda_j \xi_j, \, j = 1, 2; 0<\lambda_1\leq\lambda_2,	\\
	|\xi_j|&=1, \, \xi_2 = \Omega \, \xi_1, \, \Omega=
	\begin{pmatrix}
	0 & -1\\ 1 & 0
	\end{pmatrix}.		
\end{align} 
A particle initially located at $\mathbf x_0$ at $t_0$, when advected, moves to $F_{t_0}^{t_0+\tau}$ after a time interval $\tau$ and the amount of stretching about this trajectory can be characterized by the finite time Lyapunov exponent (FTLE) defined as
\begin{equation}
\label{eq:FTLE-def}
FTLE(\mathbf x_0, t_0) = \frac{1}{|\tau|}\ln \sqrt{\lambda_{\max}(C_{t_0}^{t_0+\tau})},
\end{equation}
where $\lambda_{\max}$ is the largest eigenvalue of $C_{t_0}^{t_0+\tau}$. In other words, it describes a finite time average of the maximum expansion rate for a pair of particles advected by the flow \cite{shadden2005definition}. 

The FTLE field is both space and time dependent and ridges of local maxima in the field represent material lines in the flow and reveal transport barriers between the different regions of the flow. 
%Thus, the FTLE field, which is both space and time-dependent, captures complex flow patterns such as vortex formation, their development. 
In \eqref{eq:FTLE-def}, an absolute value of $\tau$ is used since the FTLE can be computed for forward times ($\tau>0$) and backward times ($\tau<0$). If the FTLE field is calculated by integrating trajectories in backward time, the fluid particles tend to collect or accumulate in local structures and the corresponding ridges are called attracting (unstable) material lines. On the other hand, for the forward time, ridges correspond to the repelling (stable) lines of maximum spreading so that particles initially close diverge quickly in forward time. These separatrices are called Lagrangian Coherent Structures (LCS) and are analogue of stable and unstable manifolds in the time-dependent systems \cite{haller2015lagrangian}. 
%The stable (repelling) material lines are defined as the lines of maximum spreading so that particles initially close diverge quickly in forward time ($\tau >0$) and similarly, unstable (attracting) material lines as the ones corresponding to the maximum accumulation backwards $(\tau<0)$ in time. 
%\textit{These two types are \textcolor{red}{classified as} hyperbolic LCS and are different from the elliptic LCS which are closed material lines existing as coherent Lagrangian vortex boundaries}. 
These structures divide dynamically distinct regions in the flow and reveal geometry which is often hidden when viewing the vector field or even trajectories of the system \cite{haller2001distinguished}.

To calculate FTLE numerically, the deformation gradient tensor, needs to be estimated and to obtain that, fluid particles path positions are extracted over a finite time interval from CFD simulations described in the last section. Following the procedure \cite{shadden2005definition}, particle trajectories $\mathbf{x}(t)$ are determined numerically by solving \eqref{eq:xprime} using the Runge--Kutta (4,5) method. This is done using the MATLAB ode45 solver with an absolute integration tolerance of $10^{-6}$, for the particles to get their trajectories and final positions at time $t_0+\tau$. To obtain the velocity along particle trajectories, a linear interpolation scheme is employed in space and time and a free slip boundary condition is used and the calculation of the trajectory of particles is stopped once they exit the flow domain.  
\begin{figure}[h]
	\centering
	\begin{subfigure}[b]{0.48\textwidth}
		\centering
		\includegraphics[width=\textwidth]{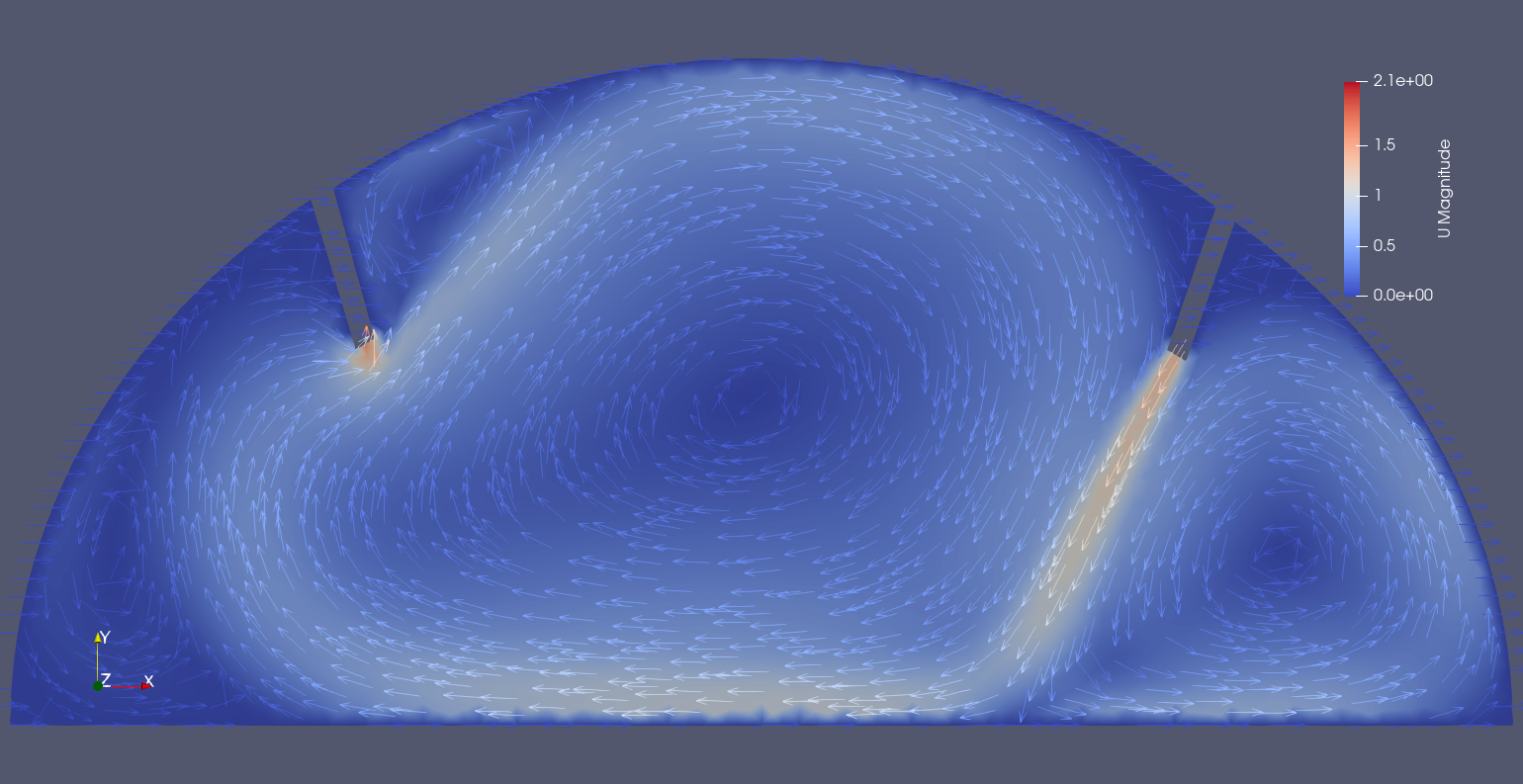}
		\caption{}
		\label{fig:VelFieldA}		
	\end{subfigure}
	\begin{subfigure}[b]{0.48\textwidth}
		\centering
		\includegraphics[width=\textwidth]{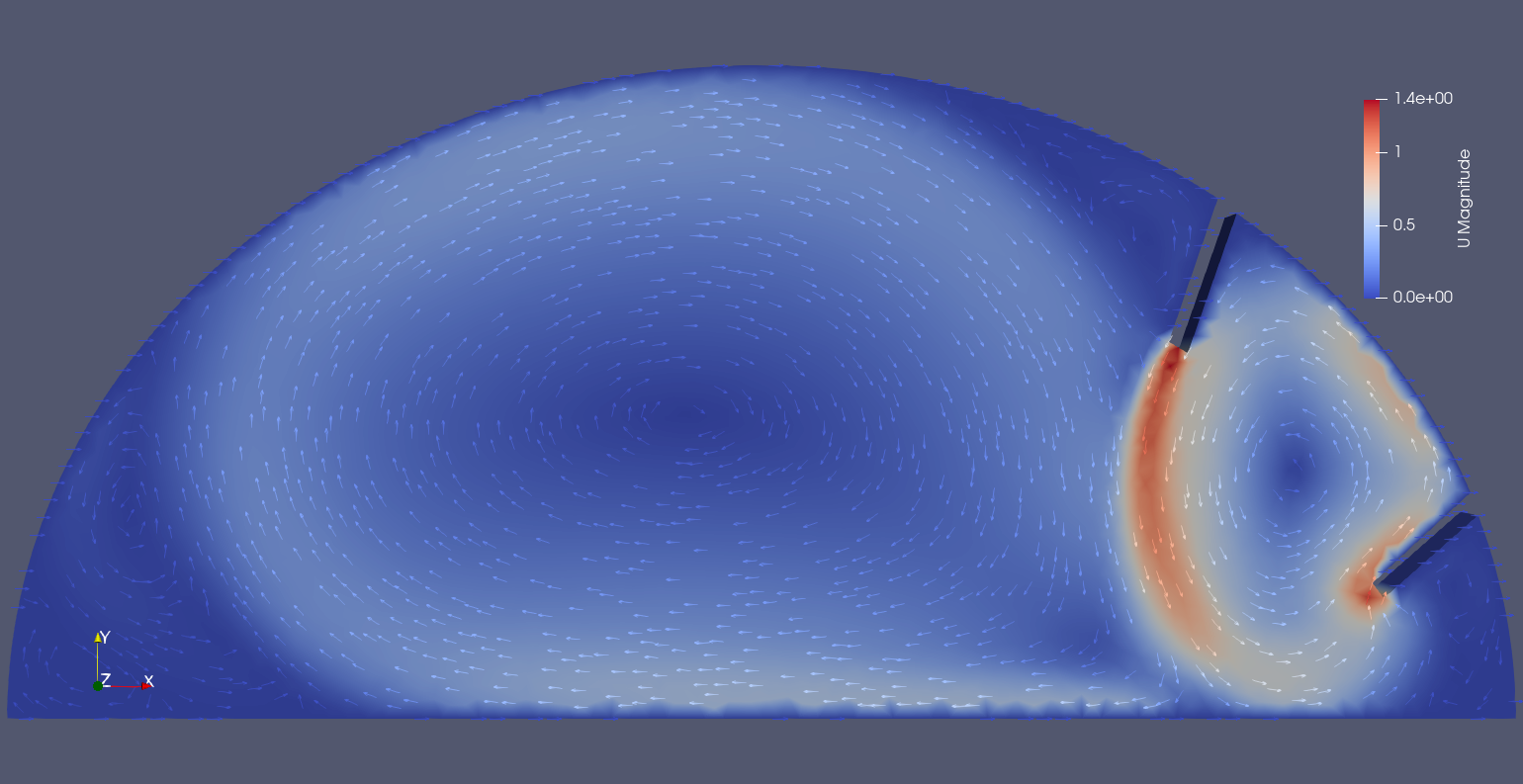}
		\caption{}
		\label{fig:VelFieldB}		
	\end{subfigure}
	\caption{The velocity field marked with arrows for two different 2D configurations, where, the left Subfigure \ref{fig:VelFieldA}, corresponds to that observed in the YZ-plane of Subfigures \ref{fig:Case1}, \ref{fig:Case3}, \ref{fig:Case4}; and the right, as in that of Subfigure \ref{fig:Case2}, where the position of the outlet port is behind the inlet/camera port. Different color profile represents the magnitude of the velocity calculated at time $t=10$ seconds.}
	\label{fig:VelField}
\end{figure}
% ------------------------------------------------------------

Thus, to implement the LCS detection in the abdomen geometry, the two-dimensional domain corresponds to the YZ-plane in Figure \ref{fig:MeshAndExpSetup}. The plane X$=0$, corresponds to the part of the abdomen that contains the `c' port, i.e., inlet and the epigastric port, i.e., outlet. The two-dimensional mesh is generated using \texttt{Gmsh v4.4.1}, and with parameters and the CFD model proposed in the last section, we calculate the velocity vector field as shown in Figure \ref{fig:VelField}. Here, the configuration of ports in Subfigure \ref{fig:VelFieldA}, corresponds to that observed in the YZ-plane of Subfigures \ref{fig:Case1}, \ref{fig:Case3}, \ref{fig:Case4}; and Subfigure \ref{fig:VelFieldB}, as in that of Subfigure \ref{fig:Case2}, where the position of the outlet port is behind the inlet/camera port. As we noticed in the last section, this dynamics is dependent on the number, position of the outlets, thus, after denoting the setup in Figure \ref{fig:VelFieldA} as the base-case, we have modified it by changing the angle of the inlet, outlet, adding one more outlet, and changing its position as in Subfigure \ref{fig:Case2}.

%\textcolor{blue}{Talk about spatial resolution.
%In order to implement the aforementioned procedure of calculating LCS lines numerically, there have been several approaches developed that mainly compute the FTLE field and hyperbolic LCS \cite{du2010transport,OHH}. }

\begin{figure}
	\centering
	\begin{subfigure}[b]{0.3\textwidth}
		\centering
		\includegraphics[width=\textwidth]{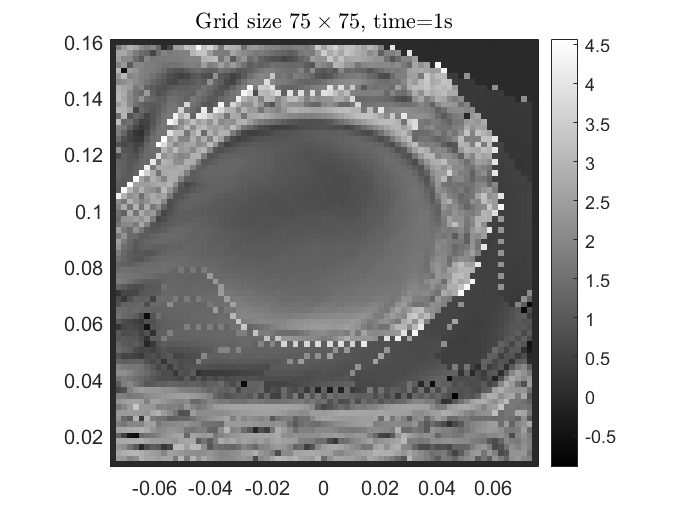}
		\caption{}
		\label{fig:BackFTLENx90t1}
	\end{subfigure}
	% 	\hfill
	\begin{subfigure}[b]{0.3\textwidth}
		\centering
		\includegraphics[width=\textwidth]{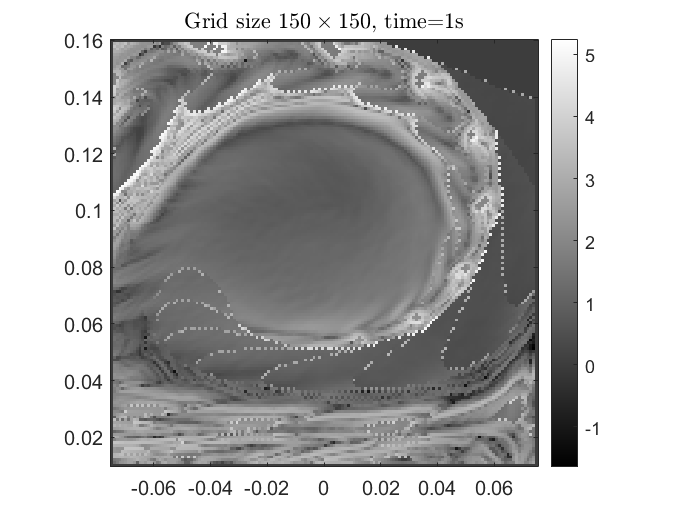}
		\caption{}
		\label{fig:BackFTLENx180t1}
	\end{subfigure}
	%  	\hfill
	\begin{subfigure}[b]{0.3\textwidth}
		\centering
		\includegraphics[width=\textwidth]{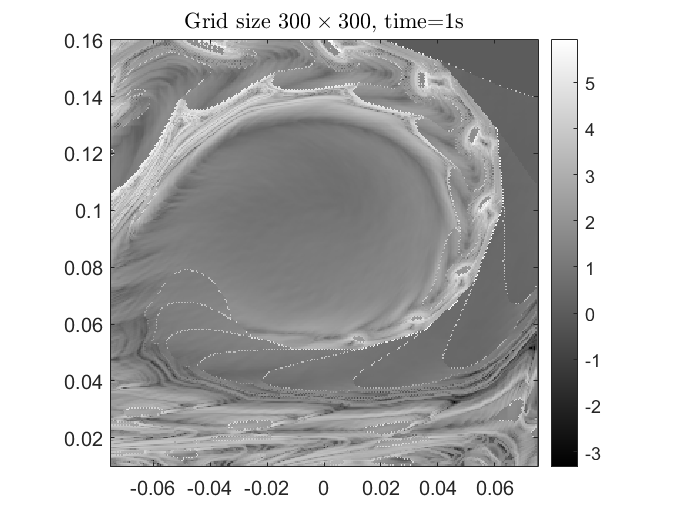}
		\caption{}
		\label{fig:BackFTLENx360t1}
	\end{subfigure}
	% 	\hfill
	\begin{subfigure}[b]{0.46\textwidth}
		\centering
		\includegraphics[width=\textwidth]{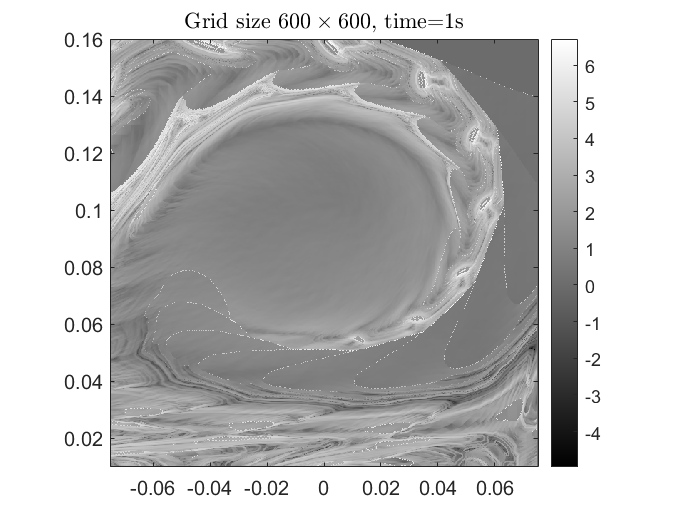}
		\caption{}
		\label{fig:BackFTLENx720t1}
	\end{subfigure}
	\begin{subfigure}[b]{0.46\textwidth}
		\centering
		\includegraphics[width=\textwidth]{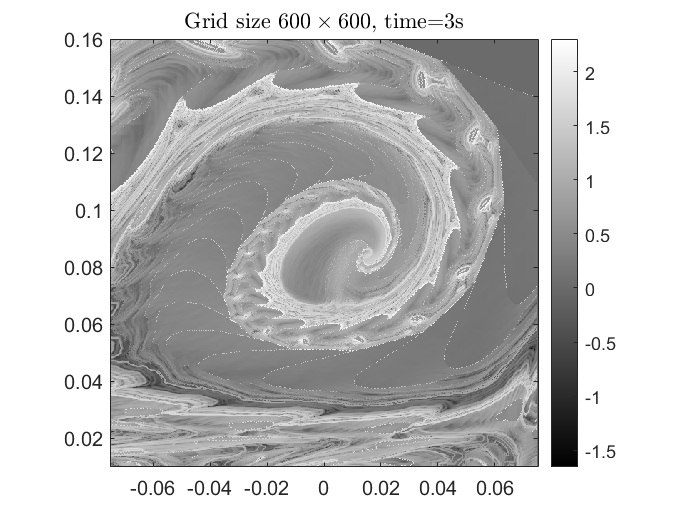}
		\caption{}
		\label{fig:BackFTLENx720t4}
	\end{subfigure}
	\caption{Backward FTLE field for a region corresponding to the central part of the geometry in Subfigure \ref{fig:VelFieldA}, with different resolutions, i.e., $75\times 75, 150\times 150, 300\times 300, 600\times 600$ for time period equal to 1s and 3s as indicated. With a finer grid and longer time period, more details about the flow are revealed.}
	\label{fig:BackFTLEdiffresolution}
\end{figure}

At first, we have considered the central part of the abdomen as in Subfigure \ref{fig:VelFieldA}. To be precise, the domain $[-0.075, 075]\times[0.01, 0.16]$, and with the velocity field obtained using the CFD simulations, we have calculated the backward FTLE field with the mesh resolution $75\times 75$. Furthermore, the integration time $t\in[4,5]$ with $\tau=1s$ was chosen to best reveal the flow structure. 
%In order to plot the FTLE values, we have first normalised them and then considered only the ones above a threshold value 0.75 setting the rest to be equal to zero. Note that changing the threshold value does not change the location of the ridges but only their width. 
With the FTLE field thus obtained in Figure \ref{fig:BackFTLEdiffresolution} the rotating pattern can be observed from the spiraling form of the FTLE field where the white color indicate its highest values and the black color its the lowest. We note that by increasing the mesh resolution, the sharpness of the FTLE field can be improved, for example, in Subfigures \ref{fig:BackFTLENx180t1}, \ref{fig:BackFTLENx360t1}, \ref{fig:BackFTLENx720t1}, we have reduced the step size $\Delta x, \Delta y$ by a factor of 2, 4 and 8; similarly, by taking longer integration time $\tau$, the flow structure was revealed more, as observed in Subfigure \ref{fig:BackFTLENx720t4}, where $\tau=3s$.

%\textcolor{red}{The hyperbolic LCS are employed for tracking debris behaviour in large complex systems but are less useful here \cite{suara2020material}. On the other hand, the elliptic LCS can indicate regions where placement of smoke evacuation may be ineffective. This is important because the objective of this work is to inform effective positioning of smoke evacuation instruments.}

%\textit{We employ \textit{LCS tool} introduced in \cite{OHH}, which can compute the elliptic LCS lines as parameterized material curves. The tool is a library of MATLAB functions that retrieves LCSs from two-dimensional unsteady flows.} Basically, it involves obtaining the particle trajectories by integrating \eqref{eq:xprime} using the Runge--Kutta (4,5) method through the MATLAB ode45 solver with an absolute integration tolerance, e.g., $10^{-6}$. Both analytical and discrete velocity fields can be used, and for the latter (as in our case), to define the right-hand side of \eqref{eq:xprime}, a linear or spline interpolation is performed both in space and time. 
%Note that, by definition the elliptic LCS lines are orthogonal and tangent to the two type of hyperbolic LCS. 

%\textit{For each of these five cases, we perform the LCS detection using the \textit{LCS tool} and compare the dynamics \cite{OHH}. }

Next, to analyze the flow in the whole abdomen domain, different rectangular sections are considered and we present the results for $[-0.18, 0.18]\times [0.01, 0.16]$, i.e., a subset of the original domain with a resolution of the main computational grid corresponding to a step size $\Delta x=\Delta y=10^{-3}$, which is much higher than the resolution of the velocity field and following the dicussion above, the integration time interval $\tau$ is chosen is chosen as 3 seconds, i.e., $t\in[4, 7]$. 

%\textit{and the $\lambda$-values are varied over the interval $[0.8, 1.2]$ with a step size $0.01$. }

\begin{figure}
	\centering
	\begin{subfigure}[b]{0.49\linewidth}
		\centering
		\includegraphics[width=\linewidth]{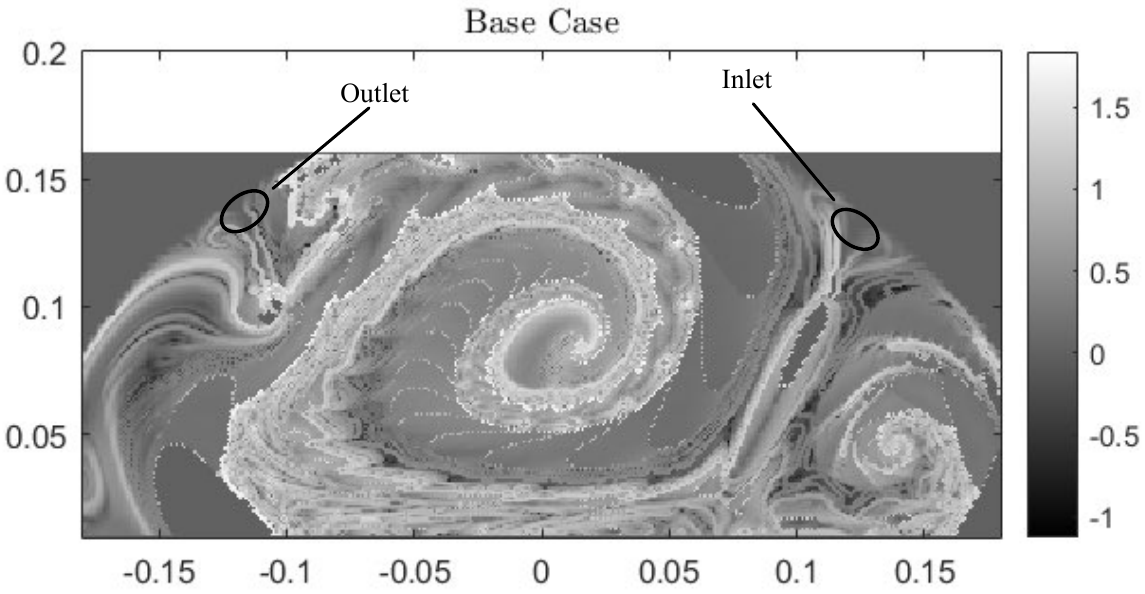}
		\caption{}
		\label{fig:BackFTLEBaseCase}
	\end{subfigure}
	% 	\hfill
	\begin{subfigure}[b]{0.49\linewidth}
		\centering	
		\includegraphics[width=\linewidth]{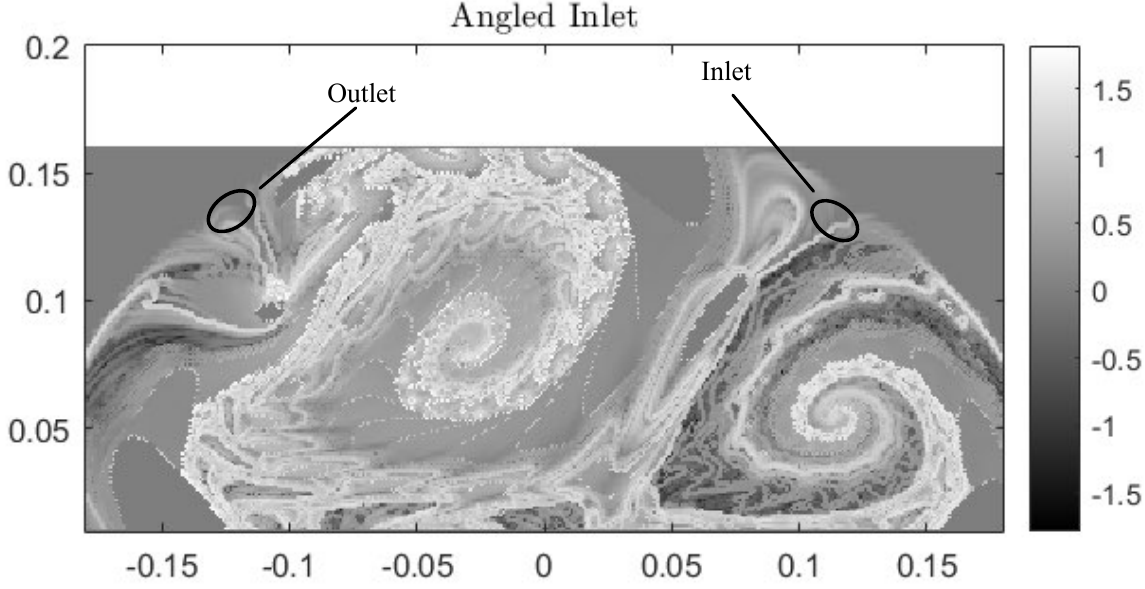}
		\caption{}
		\label{fig:BackFTLEAngledInlet}
	\end{subfigure}
	%  	\hfill
	\begin{subfigure}[b]{0.49\linewidth}
		\centering
		\includegraphics[width=\linewidth]{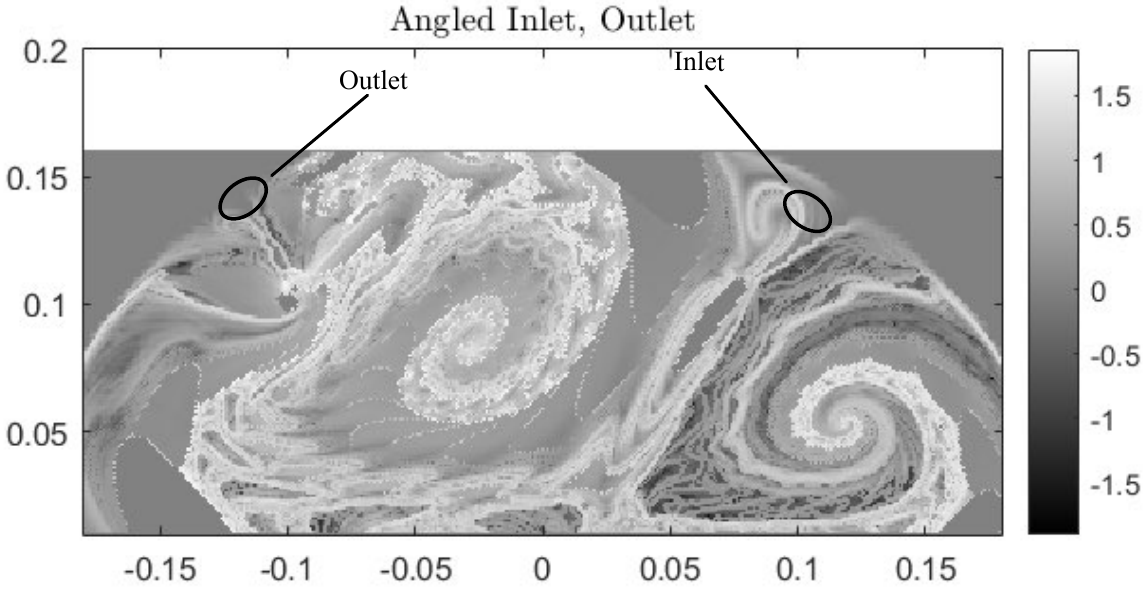}
		\caption{}
		\label{fig:BackFTLEAngledInletOutlet}
	\end{subfigure}
	% 	\hfill
	\begin{subfigure}[b]{0.49\linewidth}
		\centering
		\includegraphics[width=\linewidth]{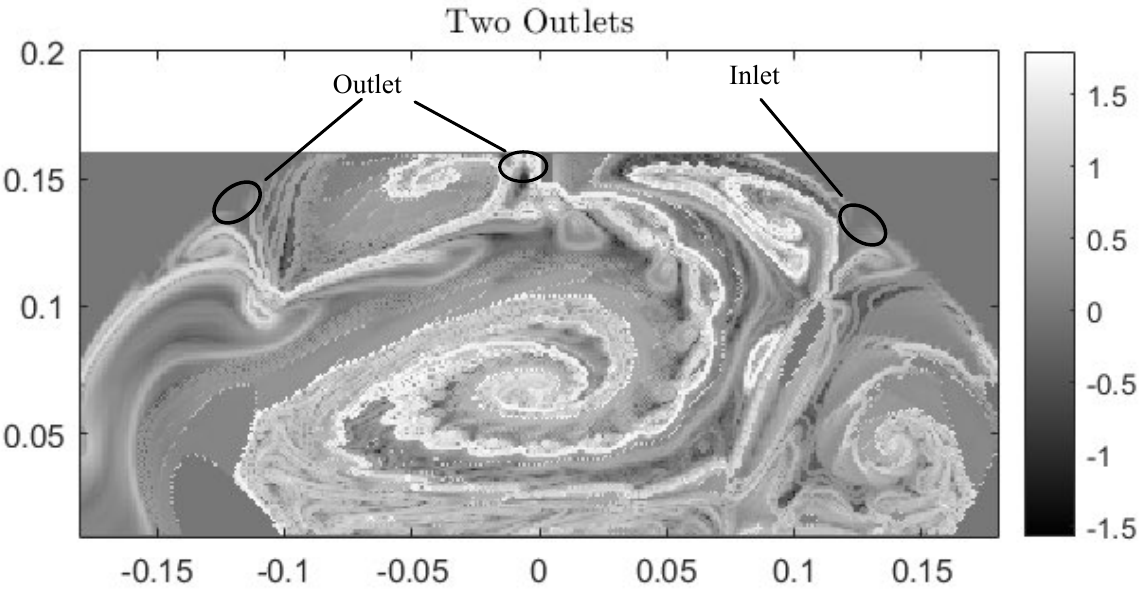}
		\caption{}
		\label{fig:BackFTLETwoOutlet}
	\end{subfigure}
	\begin{subfigure}[b]{0.49\linewidth}
		\centering
		\includegraphics[width=\linewidth]{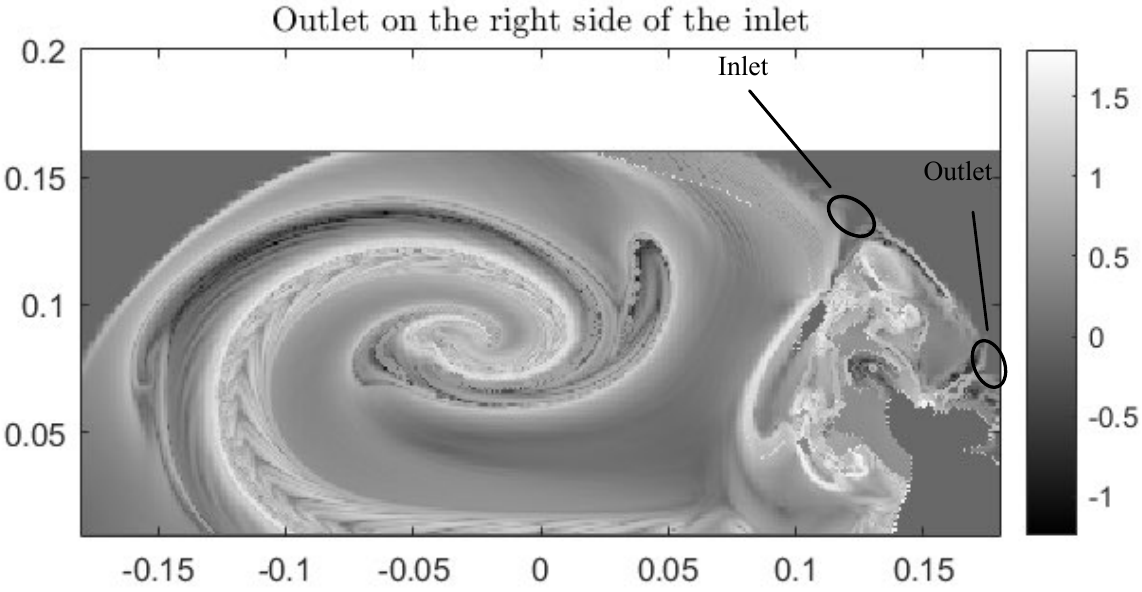}
		\caption{}
		\label{fig:BackFTLEOutletright}
\end{subfigure}
	\caption{Backward FTLE field for different geometries for time interval $[4, 7]$. The white color represents highest FTLE values, while the black color the lowest with the position of inlet and outlets indicated. Clearly, the shape of material lines are dependent on the configuration of the inlet and outlets.}
	\label{fig:BackFTLEWholeGeo}
\end{figure}

As in Section \ref{sec:MathCompModel}, we have considered different configurations by varying setups that appear in Figure \ref{fig:VelField}, for example, Figure \ref{fig:BackFTLEWholeGeo} shows the FTLE field for the base-case in Subfigure \ref{fig:BackFTLEBaseCase} and in Subfigure \ref{fig:BackFTLEAngledInlet}, the angle of the inlet is changed while in Subfigure \ref{fig:BackFTLEAngledInletOutlet}, the angle for both inlet and outlet is changed. In Subfigure \ref{fig:BackFTLETwoOutlet}, two outlets are introduced to the base-case and finally, in Subfigure \ref{fig:BackFTLEOutletright}, the outlet is placed on the right-hand side of the inlet.
% As mentioned previously, the FTLE values are normalized and we have plotted the values above the threshold value 0.71, setting the rest equal to zero. 
	In each of these five settings, from the high value of the FTLE, the positions of circular movement of the flow can be easily identified. In these plots, we have calculated the backward FTLE field where the ridges highlight the attracting/unstable material lines representing the parts of maximum accumulation. In other words, the backward FTLE trajectory integration allows to identify attractors in the flow, therefore, renders the areas of accumulation. Thus, we notice that in Subfigure \ref{fig:BackFTLEBaseCase}, there are two circulation regions, one in the region below the inlet and the another in the central area. Upon changing the angle of the inlet, in the part under the inlet, the size of material lines grow, and reduce in the central part as shown in Subfigure  \ref{fig:BackFTLEAngledInlet}. A similar behaviour is observed upon changing the outlet angle as in Subfigure  \ref{fig:BackFTLEAngledInletOutlet}. On the other hand, if we introduce one more outlet to the base case, then besides the shapes of the material lines, their number has also increased as shown in Subfigure \ref{fig:BackFTLETwoOutlet}, this implies that there are more regions of potential accumulation which, in this case, are around the two sides of the new outlet. Finally, upon moving the outlet from the left side to the right side of the inlet also changes the dynamics, i.e., the vortex formed in the center is larger and several smaller ones in the region enclosed between the inlet and the outlet are formed. Thus, we can conclude that the different configurations influence the internal dynamics of the flow; moreover, depending on the type of surgery, as there is a continuous change in the angles of inlet, outlets, given the region, the amount of accumulation would vary as well. To comment further on this in a quantitative manner and to understand the dynamics as a function of time, we characterize the overall clustering potential by calculating a spatial average of the FTLE field in different sections of the abdomen domain. This approach was proposed in \cite{dO2004} and later employed in \cite{giudici2021tracking}, where it was shown that the temporal variation of such spatially averaged positive FTLE feature measure the overall mixing strength in a system, and would quantify the overall clustering effect relative to other time instants. The computations are done using backward FTLE field as we are interested in the accumulation occurred in different part of the abdomen domain. 
	
\begin{figure}
		\centering
		\begin{subfigure}[b]{0.48\textwidth}
			\centering
			\includegraphics[width=\textwidth]{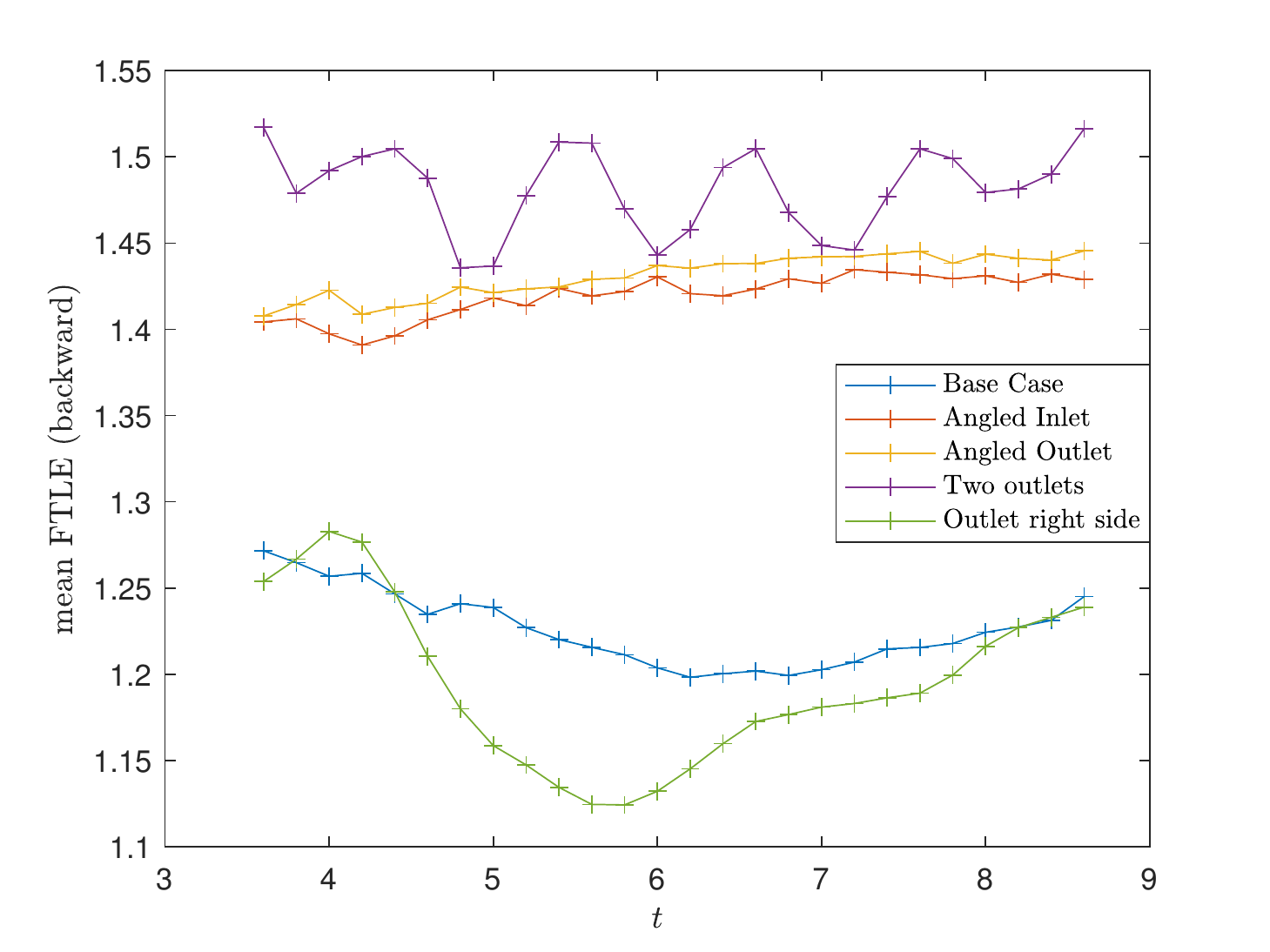}
			\caption{}
			\label{fig:ClusteringPotenRight}
		\end{subfigure}
		% 	\hfill
		\begin{subfigure}[b]{0.48\textwidth}
			\centering
			\includegraphics[width=\textwidth]{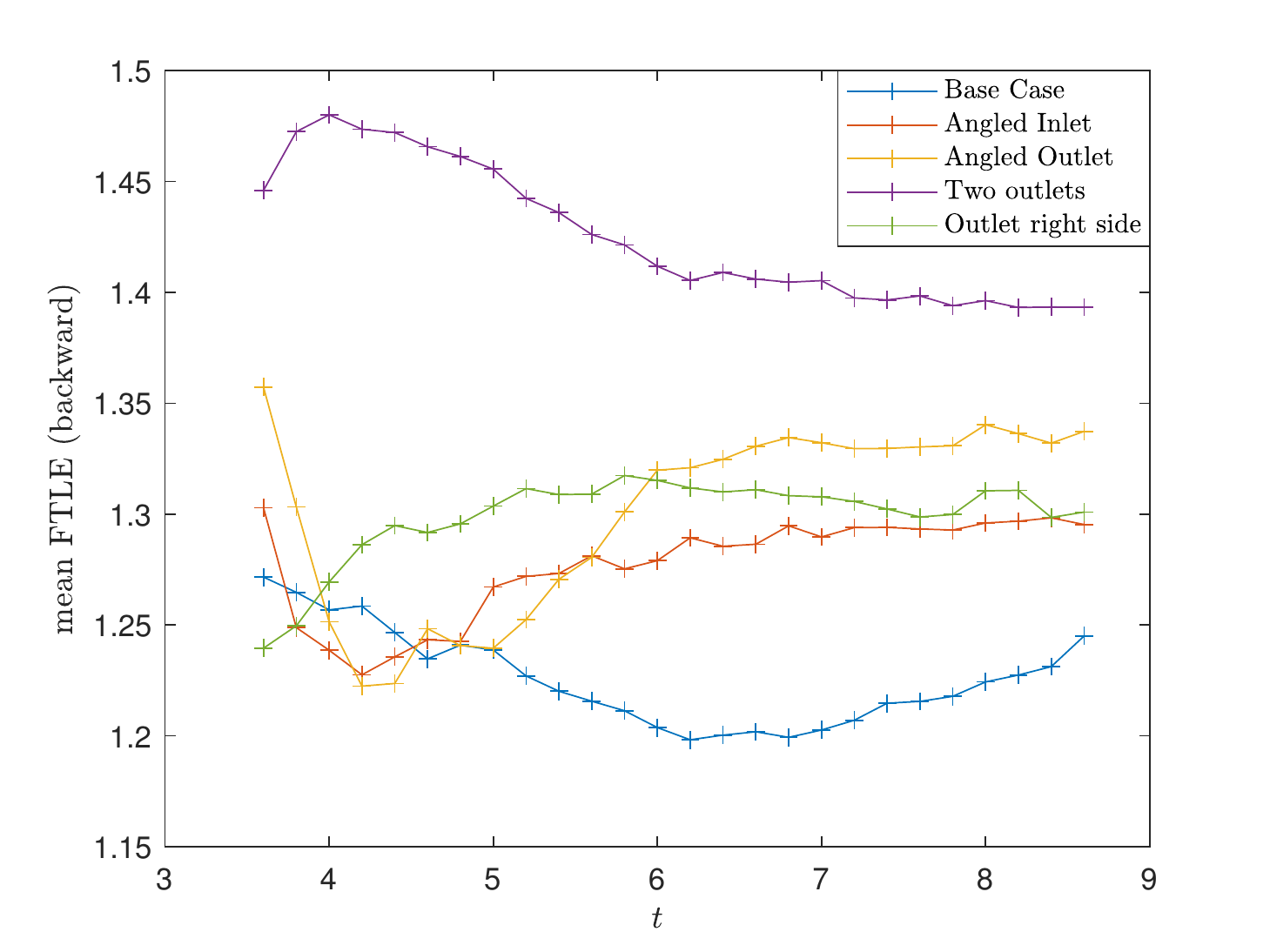}
			\caption{}
			\label{fig:ClusteringPotenCentral}
		\end{subfigure}
		\caption{Mean FTLE field for different sections of the abdomen. Left: The central part $[-0.18, 0.09]\times [0.01, 0.16]$; Right: The part under the inlet $[0.09, 0.18]\times [0.01, 0.16]$ as in Figure \ref{fig:BackFTLEWholeGeo}. Each color stands for a different case, i.e., base case, angled inlet, angled inlet and outlet, and two outlets. The increasing magnitude in the value of mean FTLE field implies that the mixing strength of the given region increases and shows a temporal variation and a comparison among different configurations.}
		\label{fig:ClusteringPotenComparison}
	\end{figure}

	In Figure \ref{fig:ClusteringPotenComparison}, we have plotted the clustering potential for different parts of the abdomen for each of the five cases for the time period $[3.6, 8.6]$. The left hand side of the figure is the clustering potential plotted for the central part of the domain, i.e., the region $[-0.18,0.09]\times [0.01, 0.16]$ and the one on the right corresponds to the region $[0.09,0.18]\times [0.09, 0.18]\times [0.01,0.16]$. We observe that the clustering potential indeed varies with the configuration of the inlets and outlets which has been indicated using different colors. This implies  that the mixing strength of the system increases as the configuration of inlet and outlets are changed. It also allows us to say that when the outlet is placed on the right side of the inlet as in Case 5, the region under the inlet will have more accumulation properties, and thus, a configuration leading to be a less optimum one; recall that Case 5 represent the YZ plane of the Case 2 of Figure \ref{fig:Exp-set-up}. A similar behaviour is observed in the central part of the abdomen as well, as shown in the left part of the Figure. 
	
%\textit{Recall that, the detection of the elliptic LCS, requires a calculation of the Poincar\'e map and to compute that, the location of the centres in the FTLE field is essential. With this, for different $\lambda$ values, closed LCS curves are calculated and the ones shown in green lines in Figure \ref{fig:FTLEAbdomenCases} correspond to the outermost marking the boundaries. In the base-case, there are two lines formed in the regions under the inlet and the large abdomen area, corresponding to the $\lambda$ values of 1.03 and 0.99, respectively.}
 
%\textit{Recall that, the Lagrangian vortex boundaries that we seek are the outermost limit cycles of the differential equation \eqref{eq:lamdalines}. In \cite{haller2013coherent}, based on the value of $\lambda$, these closed $\lambda$-lines were classified as the one with boundary weakly stretching $(\lambda>1)$, or weakly contracting $(\lambda>1)$, or non-stretching $(\lambda=1)$. Due to its properties such as conservation of arc length and that of the enclosed area in the incompressible domain, a non-stretching closed line generates strong coherence patterns. On the other hand, weakly compressing Lagrangian vortices evolve in time to gain fuller coherence while preserving their enclosed area and in this process, mild bumps in their boundaries are smoothed out while weakly stretching Lagrangian vortices can be seen as regions of gradually decreasing coherence where a non-stretching core will not exist. }

It is of interest to remove smoke before it obscures visualisation and to limit patient exposure and we see that through LCS we can identify the regions inside the abdomen where the maximum accumulation caused by the surgical smoke can take place. These findings on the FTLE field clearly captures the inflow jet impingement and the resulting vortex behaviour in the flow which is responsible for the transport, mixing and accumulation of the material. While the flow will be 3D in nature there is a bias in the superior direction. As observed through Figure \ref{fig:Exp-set-up}, three of the four cases locate trocars where effective smoke evacuation may be achieved while Case 2, i.e., Subfigure \ref{fig:Case2} would be expected to be have reduced performance. This has confirmed also through our computations on the clustering potential which is larger when compared with the base case throughout the time interval.

It is worth noting that the scenario that we have considered for the LCS analysis is rather simplified, i.e., we have considered a two-dimensional geometry with a flat surface; however, the observations that we have made are rather striking and imply a great potential in the applicability of LCS in understanding the three dimensional flow as well. Needless to say, in that case, the problem will also be computationally expensive.

%\textcolor{red}{The more complex dynamics would suggest that smoke evacuation placed in the secondary vortex would be less effective as material transport from the primary vortex (wherein smoke would be generated during a procedure) will need to cross to the secondary vortex. In general, the primary vortex appears to encompass the work area where instruments generate smoke. Placement of smoke removal technologies in this space would therefore be expected to be effective. The use of more than one smoke evacuation sink would suggest that this is less effective.}

%\textcolor{red}{It is of interest to remove smoke before it obscures visualisation and to limit patient exposure. The FTLE field clearly captures the inflow jet impingement and the resulting vortex behaviour. While the flow will be 3D in nature there is a bias in the superior direction. In Figure \ref{fig:Exp-set-up}, three of the four cases locate trocars where effective smoke evacuation may be achieved while Case 2, i.e., Subfigure \ref{fig:Case2} would be expected to be have reduced performance.}

The coherent structures detected are responsible for the Lagrangian transport (the movement and mixing) in the flow as found for garbage, oil contamination in ocean flow which can be compared with that of the surgical smoke in case of laparoscopic surgery. In this work, we have successfully proved their applicability for this problem by identifying area of maximum accumulation. It is desirable to extend this approach to a three-dimensional flow with a more complex geometry of the real abdomen with a non-flat base and compare the results which we plan to address in future. In conclusion, we have provided a novel starting point to apply LCS analysis in the field of medical surgery which is also capable of addressing complicated dynamics of surgical smoke inside the abdomen.

\section*{Acknowledgments}
This work has been supported by the project PORSAV (Protecting Operating Room Staff Against Viruses) funded by the European Union’s Horizon 2020 research and innovation programme under grant agreement No 101015941. The authors wish to acknowledge the Irish Centre for High-End Computing (ICHEC) and the ResearchIT Sonic cluster for the provision of computational facilities and support. 

\bibliography{LCSLaparoscopy}

\begin{thebibliography}{10}

\bibitem{sauerland2006laparoscopy}
S.~Sauerland, F.~Agresta, R.~Bergamaschi, G.~Borzellino, A.~Budzynski,
  G.~Champault, A.~Fingerhut, A.~Isla, M.~Johansson, P.~Lundorff, {\em et~al.},
  ``Laparoscopy for abdominal emergencies,'' {\em Surgical Endoscopy and Other
  Interventional Techniques}, vol.~20, no.~1, pp.~14--29, 2006.

\bibitem{McCauley2010}
G.~McCauley, ``Understanding electrosurgery,'' {\em Bovie Med Corp}, vol.~4,
  pp.~4--15, 2010.

\bibitem{FCC}
J.~K.-M. Fan, F.~S.-Y. Chan, and K.-M. Chu, ``Surgical smoke,'' {\em Asian
  Journal of Surgery}, vol.~32, no.~4, pp.~253--257, 2009.

\bibitem{dalli2020gas}
J.~Dalli, M.~F. Khan, K.~Nolan, and R.~A. Cahill, ``Gas leaks through
  laparoscopic energy devices and robotic instrumentation-video vignette,''
  {\em Colorectal Disease}, 2020.

\bibitem{dalli2020laparoscopic}
J.~Dalli, M.~F. Khan, K.~Nolan, and R.~A. Cahill, ``Laparoscopic
  pneumoperitoneum escape and contamination during surgery using the airseal
  insufflation system: video vignette,'' {\em Colorectal Disease}, 2020.

\bibitem{hardy2021aerosols}
N.~Hardy, J.~Dalli, M.~Khan, K.~Nolan, and R.~Cahill, ``Aerosols, airflow, and
  airspace contamination during laparoscopy,'' {\em British Journal of
  Surgery}, vol.~108, no.~9, pp.~1022--1025, 2021.

\bibitem{mac2022aerosol}
M.~Mac Giolla~Eain, R.~Cahill, R.~MacLoughlin, and K.~Nolan, ``Aerosol release,
  distribution, and prevention during aerosol therapy: a simulated model for
  infection control,'' {\em Drug Delivery}, vol.~29, no.~1, pp.~10--17, 2022.

\bibitem{CDKFN}
R.~Cahill, J.~Dalli, M.~Khan, M.~Flood, and K.~Nolan, ``Solving the problems of
  gas leakage at laparoscopy,'' {\em The British journal of surgery}, 2020.

\bibitem{ZBF}
M.~H. Zheng, L.~Boni, and A.~Fingerhut, ``Minimally invasive surgery and the
  novel coronavirus outbreak: lessons learned in china and italy,'' {\em Annals
  of surgery}, 2020.

\bibitem{crowley2022cfd}
C.~Crowley, R.~Cahill, and K.~Nolan, ``A cfd analysis of gas leaks and aerosol
  transport in laparoscopic surgery,'' {\em Physics of Fluids}, vol.~34, no.~8,
  p.~081905, 2022.

\bibitem{CY}
W.~K. Chow and R.~Yin, ``A new model on simulating smoke transport with
  computational fluid dynamics,'' {\em Building and Environment}, vol.~39,
  no.~6, pp.~611--620, 2004.

\bibitem{NSCTC}
H.~H. Najafabadi, V.~Suresh, C.~J.~T. Spence, E.-L. Teh, and J.~E. Cater,
  ``Development of a numerical model of surgical smoke during laparoscopy,''
  {\em International Journal of Heat and Mass Transfer}, vol.~175, p.~121253,
  2021.

\bibitem{shadden2005definition}
S.~C. Shadden, F.~Lekien, and J.~E. Marsden, ``Definition and properties of
  lagrangian coherent structures from finite-time lyapunov exponents in
  two-dimensional aperiodic flows,'' {\em Physica D: Nonlinear Phenomena},
  vol.~212, no.~3-4, pp.~271--304, 2005.

\bibitem{haller2015lagrangian}
G.~Haller, ``Lagrangian coherent structures,'' {\em Annual Review of Fluid
  Mechanics}, vol.~47, pp.~137--162, 2015.

\bibitem{beron2013objective}
F.~J. Beron-Vera, Y.~Wang, M.~J. Olascoaga, G.~J. Goni, and G.~Haller,
  ``Objective detection of oceanic eddies and the agulhas leakage,'' {\em
  Journal of Physical Oceanography}, vol.~43, no.~7, pp.~1426--1438, 2013.

\bibitem{vetel2009lagrangian}
J.~V{\'e}tel, A.~Garon, and D.~Pelletier, ``Lagrangian coherent structures in
  the human carotid artery bifurcation,'' {\em Experiments in fluids}, vol.~46,
  no.~6, pp.~1067--1079, 2009.

\bibitem{suara2020material}
K.~Suara, M.~Khanarmuei, A.~Ghosh, Y.~Yu, H.~Zhang, T.~Soomere, and R.~J.
  Brown, ``Material and debris transport patterns in moreton bay, australia:
  The influence of lagrangian coherent structures,'' {\em Science of The Total
  Environment}, vol.~721, p.~137715, 2020.

\bibitem{fluids1040038}
K.~May-Newman, V.~Vu, and B.~Herold, ``Modeling the link between left
  ventricular flow and thromboembolic risk using lagrangian coherent
  structures,'' {\em Fluids}, vol.~1, no.~4, 2016.

\bibitem{giudici2021tracking}
A.~Giudici, K.~A. Suara, T.~Soomere, and R.~Brown, ``Tracking areas with
  increased likelihood of surface particle aggregation in the gulf of finland:
  A first look at persistent lagrangian coherent structures (lcs),'' {\em
  Journal of Marine Systems}, vol.~217, p.~103514, 2021.

\bibitem{johnson1997laparoscopic}
A.~Johnson, ``Laparoscopic surgery,'' {\em the Lancet}, vol.~349, no.~9052,
  pp.~631--635, 1997.

\bibitem{weller1998tensorial}
H.~G. Weller, G.~Tabor, H.~Jasak, and C.~Fureby, ``A tensorial approach to
  computational continuum mechanics using object-oriented techniques,'' {\em
  Computers in physics}, vol.~12, no.~6, pp.~620--631, 1998.

\bibitem{haller2001distinguished}
G.~Haller, ``Distinguished material surfaces and coherent structures in
  three-dimensional fluid flows,'' {\em Physica D: Nonlinear Phenomena},
  vol.~149, no.~4, pp.~248--277, 2001.

\bibitem{dO2004}
F.~d'Ovidio, V.~Fern{\'a}ndez, E.~Hern{\'a}ndez-Garc{\'\i}a, and C.~L{\'o}pez,
  ``Mixing structures in the mediterranean sea from finite-size lyapunov
  exponents,'' {\em Geophysical Research Letters}, vol.~31, no.~17, 2004.

\end{thebibliography}
\bibliographystyle{ieeetr}

\end{document}